\documentclass[superscriptaddress,aps,prb,twocolumn,floatfix]{revtex4-2}

\usepackage{amssymb, amsmath}
\usepackage{graphicx}
\usepackage{bm}
\usepackage{nicefrac}
\usepackage{hyperref}
\usepackage{booktabs}
\usepackage{multirow}

\hypersetup
    {colorlinks=true,citecolor=blue,urlcolor=blue,linkcolor=blue}
	
\newcommand{\fzmo}[0]{Fe$_{1.86}$Zn$_{0.14}$Mo$_3$O$_8$}
\newcommand{\fzmox}[0]{(Fe$_{1-y}$Zn$_{y}$)$_2$Mo$_3$O$_8$}
\newcommand{\fmo}[0]{Fe$_{2}$Mo$_3$O$_8$}

\begin{document}

\title
{Magnetic and vibronic THz excitations in Zn doped \fmo}

\author{B.~Csizi}
\author{S.~Reschke}
\author{A.~Strini\'{c}}
\affiliation{Experimentalphysik V, Center for Electronic
Correlations and Magnetism, Institute for Physics, Augsburg
University, D-86135 Augsburg, Germany}
\author{L.~Prodan}
\author{V.~Tsurkan}
\affiliation{Experimentalphysik V, Center for Electronic
Correlations and Magnetism, Institute for Physics, Augsburg
University, D-86135 Augsburg, Germany} \affiliation{Institute of
Applied Physics, MD-2028~Chi\c{s}in\u{a}u, Republic of Moldova}
\author{I.~K\'ezsm\'arki}
\author{J.~Deisenhofer}
\affiliation{Experimentalphysik V, Center for Electronic
Correlations and Magnetism, Institute for Physics, Augsburg
University, D-86135 Augsburg, Germany}

\date{\today}

\begin{abstract}
We report on optical excitations in the magnetically ordered phases of multiferroic
\fzmo{} in the
frequency range from 10-130~cm$^{-1}$ (0.3-3.9\,THz). In
the collinear easy-axis antiferromagnetic phase below $T_N=50$~K eleven
optically active modes have been observed in finite magnetic fields, assuming
that the lowest-lying mode is doubly degenerate. The large number of modes
reflects either a more complex magnetic structure than in pure \fmo{} or
that spin stretching modes become active in addition
to the usual spin precessional modes. Their magnetic field dependence, for
fields applied along the easy axis, reflects the irreversible magnetic-field
driven phase transition from the antiferromagnetic ground state to a
ferrimagnetic state, while the number of modes remains unchanged in the
covered frequency region. We determined selection rules for some of
the AFM modes by investigating all polarization configurations and identified
magnetic- and electric-dipole active modes as well. In addition to these
sharp resonances, a broad electric-dipole active excitation band, which is not influenced by the external magnetic field, occurs below $T_N$ with an onset at 12~cm$^{-1}$. We are able to model
this absorption band as a vibronic excitation related to the lowest-lying
Fe$^{2+}$ electronic states in tetrahedral environment.
\end{abstract}

\maketitle

\section{Introduction}
Multiferroic compounds have been in the focus of condensed matter research for the last two decades \cite{Spaldin2019,Dong2019} and, more recently, have attracted interest due to the possible dynamic magneto-electric effects, such as, for example,  uni-directional light propagation \cite{Kezsmarki2015} and potential applications in spintronics based on antiferromagnetic materials \cite{Jungwirth2018,Baltz2018}.

\begin{figure}[t]
    \centering
    \includegraphics[width=\linewidth]{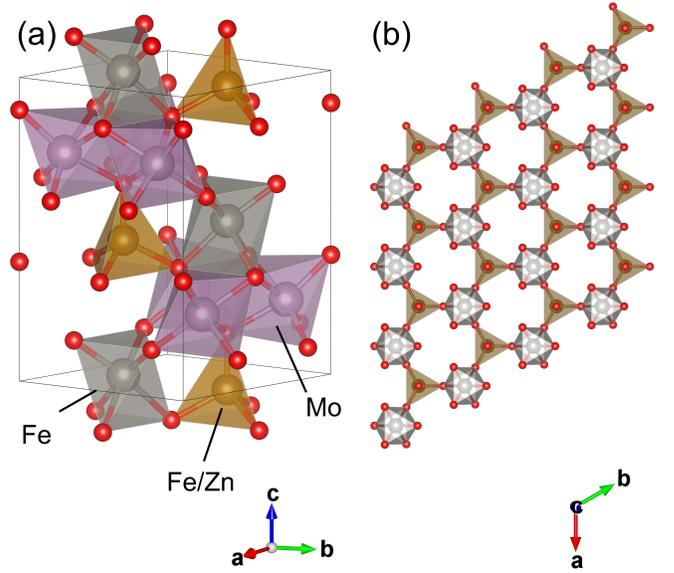}
    \caption{\label{fig:structure} (a) Double unit cell of
	\fzmox{}, (b) top view onto the honeycomb-like layers
	stacked along the $c$-axis.}
\end{figure}

The multiferroic compound \fmo{}, which is known as the mineral
kamiokite \cite{Sasaki1985}, adopts a hexagonal unit cell with lattice
constants $a=5.777$\,\AA, $c=10.057$\,\AA{} at room temperature as
depicted in Fig.~\ref{fig:structure}(a)
\cite{Bertrand1975}. It belongs to the polar space group $P6_3mc$
with a polarization along the $c$ axis \cite{McCarroll1957, McAlister1983}.
The Fe$^{2+}$ ions occupy two different sites with tetrahedral (site $A$)
and octahedral oxygen coordination (site $B$) \cite{Wang2015}. The two types
of corner-sharing polyhedra form honeycomb-like layers in the $ab$-plane
separated by breathing kagome layers of molybdenum octahedra as shown in
Fig.~\ref{fig:structure}(b). The Mo layers do not contribute to the
magnetism due to spin-singlet formation of structural
Mo$_3$O$_{13}$ clusters
\cite{Varret1972, Cotton1964}. \fmo{} exhibits a collinear
antiferromagnetic (AFM) order of the iron ions below the Néel
temperature $T_N=60\,$K,
which is accompanied by a strong increase of the polarization
\cite{Wang2015, Kurumaji2015}, giving rise to a type-I multiferroic state
and strong magneto-optical effects \cite{Kurumaji2017a, Kurumaji2017, Yu2018}.
In addition, an applied
magnetic field along the $c$ axis induces a ferrimagnetic (FiM) order \cite{Wang2015, Kurumaji2015}. Recently, the
possibility of additional orbital ordering on the Fe sites in both the
AFM and the FiM state has been suggested by ab-initio calculations
\cite{Solovyev2019}. \\
\indent Isovalent substitution of Fe by nonmagnetic Zn in \fzmox{}
has been reported to influence the magnetic phase diagram, leading
to a stabilization of the ferrimagnetic phase with increasing Zn
content \cite{Kurumaji2015, Nakayama2011}.  Previous studies
revealed that the Zn$^{2+}$ ions predominantly occupy the
tetrahedral sites for $y\leq 0.5$ \cite{Varret1972, Streltsov2019}.
Both the antiferromagnetic and the ferrimagnetic phases of \fzmox{} reportedly exhibit interesting electric- and magnetic-dipole active magnon excitations and, furthermore, intricate magneto-optical effects, e.g. gyrotropic birefringence \cite{Kurumaji2017} and the optical diode effect \cite{Yu2018}.
In this study we investigated the low-energy excitations in
single crystals of \fzmo{} by THz spectroscopy. At this Zn concentration
we have the possibility to study the dynamic fingerprints of the magnetic-field induced transition regime and both phases in a suitable magnetic-field range: For $y=0.05$ the FiM phase for $T<30$~K can only be reached for $H>5$~T and for $y=0.125$ the antiferromagnetic phase reportedly occurs only as a metastable state \cite{Kurumaji2015}. Moreover, magnetic-field dependent THz measurements in Faraday configuration have not been reported previously for $y=0$ or $y=0.125$ \cite{Kurumaji2017a}, making  a value $0.05< y< 0.125$ a suitable concentration to shed light on the dynamical properties of both phases in a detailed THz transmission study.

\section{Experimental Details}
Polycrystalline \fzmox{} samples were prepared by repeated synthesis at
1000$\,^{\circ}$C of binary oxides FeO (99.999\%), MoO$_2$ (99\%), and ZnO in
evacuated quartz ampoules aiming for a concentration with $x=0.1$. Single crystals were grown by the
chemical transport reaction method at temperatures between 950 and
900$\,^{\circ}$C. TeCl$_4$ was used as the source of the transport
agent. Large single crystals up to 5 mm have been obtained after
4 weeks of transport.
The X-ray diffraction pattern of the crushed
single crystals shown in Fig.~\ref{fig:XRD_Cp}(a) revealed a single-phase composition with a hexagonal
symmetry using space group $P6_3mc$ and a Zn content corresponding to $x\approx 0.07$. The obtained lattice constants are
$a=b=5.773(2)$\,\AA{} and $c=10.017(2)$\,\AA{}.

The specific heat
was measured in a Quantum Design physical properties measurement
system from 1.8 to 300 K and in fields up to 9~T.
Magnetization measurements were performed using a
superconducting quantum interference device magnetometer (Quantum
Design MPMS-5).

Transmission measurements in the frequency range from
10-130\,cm$^{-1}$ were performed by THz time-domain spectroscopy  using
a Toptica TeraFlash spectrometer and an Oxford Instruments Spectromag
cryomagnet in external magnetic fields ranging from -7\,T to 7\,T in an
$ac$-cut sample with a thickness of $d=1.48$\,mm and an $ab$-cut sample
with $d=0.6$\,mm.

\section{Sample Characterization and Phase Diagram}

\begin{figure}[htb]
    \centering
    \includegraphics[width=\linewidth]{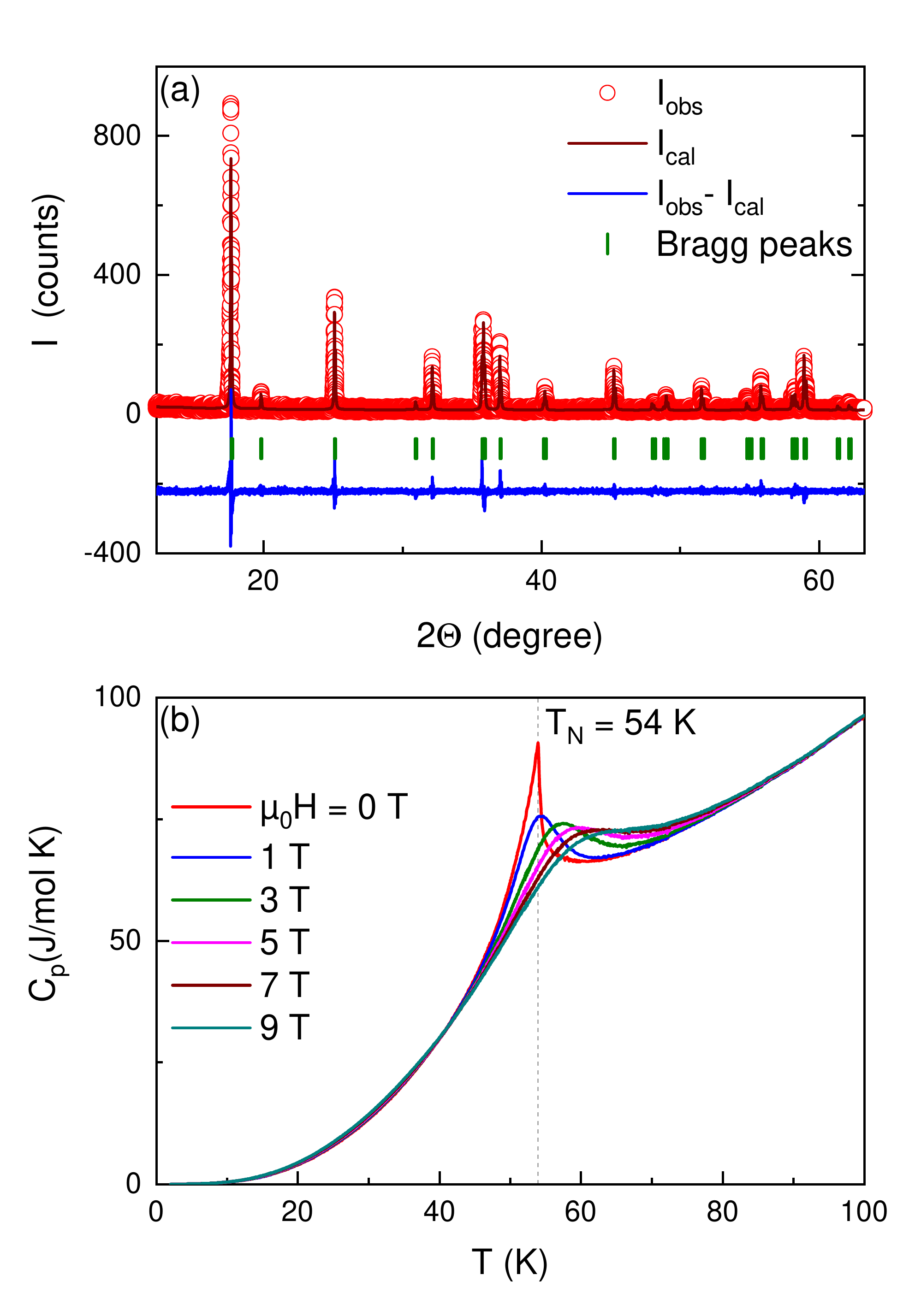}
    \caption{\label{fig:XRD_Cp} (a) Room temperature XRD spectra of \fzmo{} and the corresponding calculated pattern for space group $P6_3mc$.
    (b) Temperature dependent specific heat $C_p$ in the vicinity of the magnetic ordering transition in magnetic fields applied along the $c$-axis between 0-9~T.}
\end{figure}

The temperature dependence of the specific heat shown in Fig.~\ref{fig:XRD_Cp}(b) reveals a clear $\lambda$-like anomaly at the magnetic ordering transition $T_N=50$~K, similar to the reported anomaly of pure \fmo{} \cite{Wang2015}.  With an increasing external magnetic field applied along the crystallographic $c$-axis the anomaly broadens and shifts to higher temperatures. The transition temperatures were estimated using the maxima of the specific heat.

\begin{figure}[htb]
    \centering
    \includegraphics[width=\linewidth]{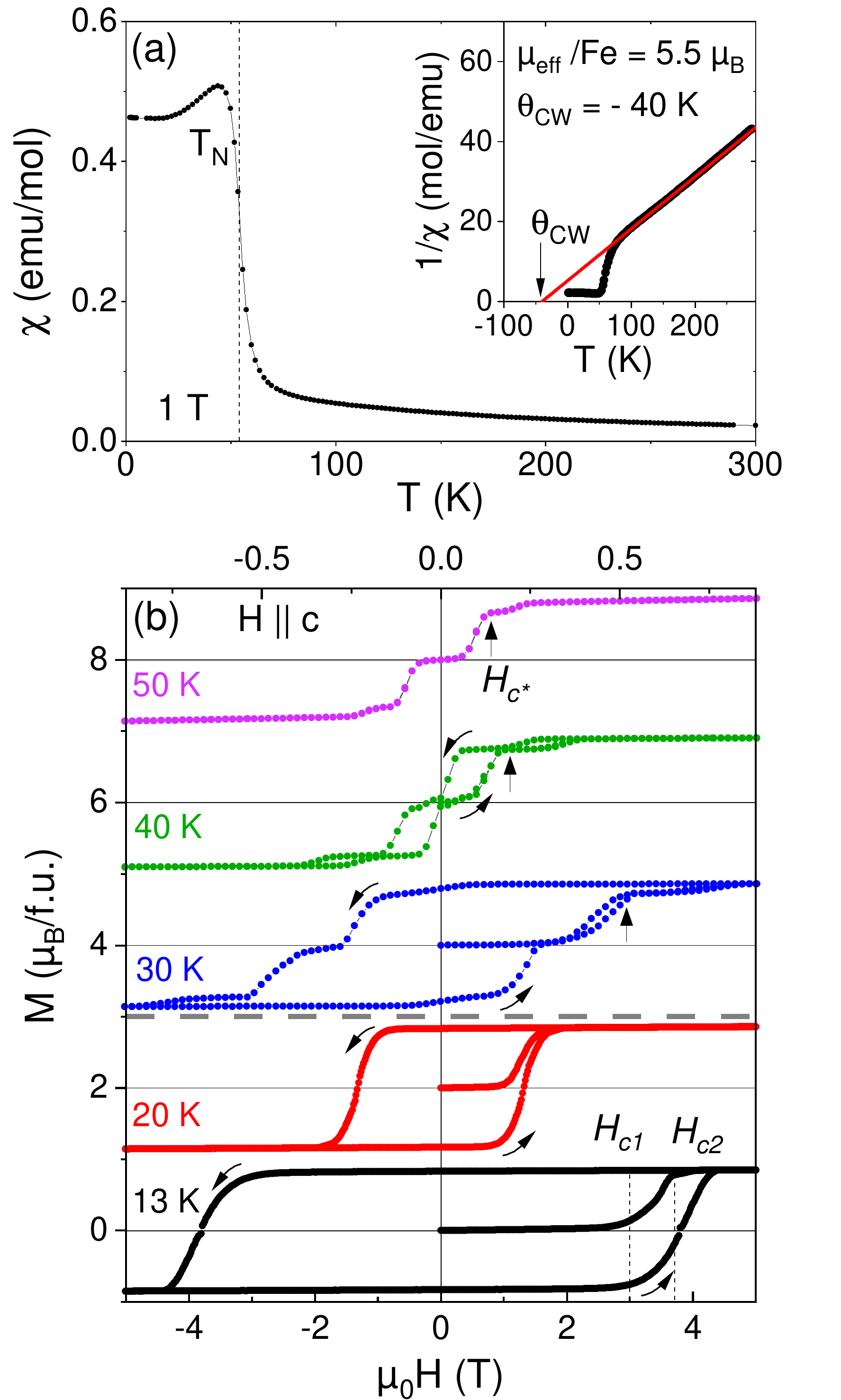}
    \caption{\label{fig:magnetization} (a) Temperature dependence of
    the magnetic susceptibility $\chi$ measured in 1\,T for
    $H\parallel c$. The inset shows the temperature dependence of
    $\chi^{-1}$ and a Curie-Weiss fit (solid red line); (b) Field
    dependent magnetization curves for several temperatures
    below $T_N$.}
\end{figure}

The temperature
dependence of the magnetic susceptibility measured in 1~T shown in
Fig.~\ref{fig:magnetization}(a) exhibits a steep increase and a subsequent maximum at the same temperature as the $\lambda$-anomaly in the specific heat at $T_N=50$~K confirming that the transition corresponds to the magnetic ordering transition in agreement with previous studies \cite{Bertrand1975,McAlister1983,Kurumaji2015}.

The temperature dependence of $1/\chi$ in the para\-magnetic state range
from 100-300\,K can be described by a Curie-Weiss law  yielding
$\Theta_{\mathrm{CW}}=-40$~K and an effective moment
$\mu_{\mathrm{eff}}=5.5\,\mu_\mathrm{B}$, which is enhanced in
comparison to the spin-only value of
$\mu_{\mathrm{eff}}=4.9\,\mu_\mathrm{B}$ expected for Fe$^{2+}$
ions with spin $S=2$, assuming that all Zn ions occupy tetrahedral
sites and that the effective $g$-factors are 2.1 \cite{Bertinshaw2014}
and 2.0 \cite{Abragam1970}
for the tetrahedral and octahedral Fe sites,
respectively. The effective magnetic moment is comparable to reported values of $\mu_{\mathrm{eff}}=5.7\,\mu_\mathrm{B}$ for pure \fmo{} \cite{Bertrand1975}. The enhancement compared to the spin-only value could be due to a sizeable orbital
contribution \cite{Solovyev2019, Reschke2020}. The obtained Curie-Weiss temperature is
clearly reduced in comparison with the values $\Theta_{\mathrm{CW}}=-110$~K or 100~K reported for \fmo{} \cite{Bertrand1975,Nakayama2011}, but in line with the reported tendency for
increasing Zn substitution, where  $\Theta_{\mathrm{CW}}$ eventually
becomes positive \cite{Nakayama2011}.

\begin{figure}[htb]
    \centering
    \includegraphics[width=\linewidth]{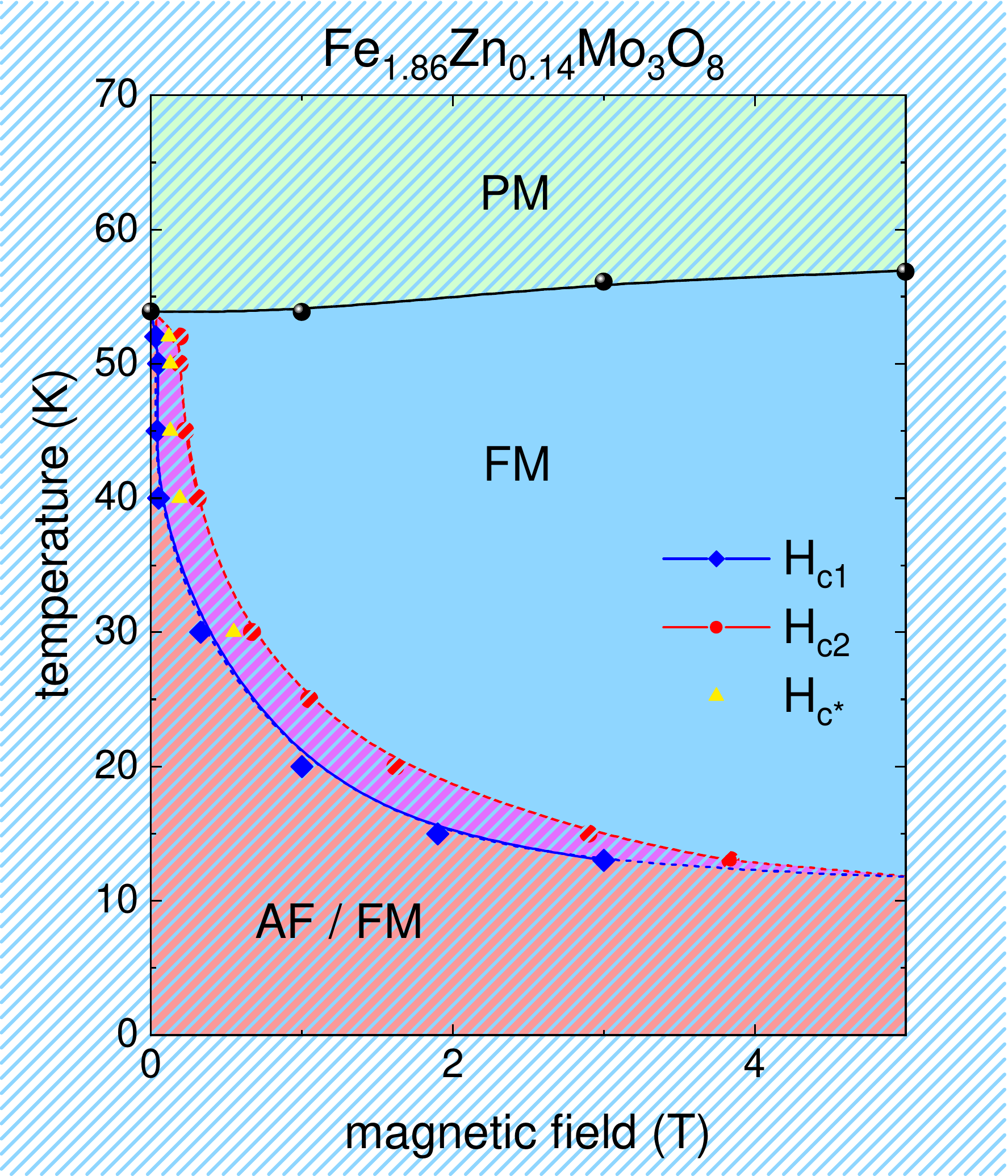}
    \caption{\label{fig:phasediagram} $H$-$T$ phase diagram of \fzmo{} established from specific heat and magnetization measurements. The dashed areas indicate the metastable ferrimagnetic phase as described in the text.}
\end{figure}

The magnetic-field dependence of the magnetization for $H\parallel c$ is shown for several temperatures below $T_N$ in Fig.~\ref{fig:magnetization}(b). At 13~K the transition from the antiferromagnetic to the ferrimagnetic state is taking place via a two-step feature characterized by the critical fields $H_{c1}=3$~T
and $H_{c2}=3.8$~T as indicated in  Fig.~\ref{fig:magnetization}(b). This is in agreement with previous studies for samples with $y=0.1$ \cite{Kurumaji2015}. However, at 50~K, just below the magnetic transition transition temperature, no hysteretic behavior and formation of a metastable state is observed, but an intermediate third step is observed at $H_{c*}$. At 40~K a hysteretic behaviour appears, but the remanent magnetization is still zero, while the width of the hysteresis is clearly increased at 30~K and the remanent magnetization already corresponds to the metastable ferrimagnetic configuration.
At 20~K the intermediate step at $H_{c*}$ is not discernible anymore and an almost symmetric hysteresis has evolved. In addition, the widths of the hysteretic curves measured by the distance of the coercive fields at 30~K and 20~K are 0.27~ and 1.3~T, respectively, seem to be enhanced in comparison with the widths shown in Ref. \onlinecite{Kurumaji2015} for $y=0.1$. This is in agreement with a less diluted system with a Zn concentration $y=0.07$.

In Fig.~\ref{fig:phasediagram} the critical fields and temperatures obtained from specific heat and magnetization measurements are summarized in a $H$-$T$-diagram, which can be compared with the phase diagrams for $y=0.05$ and $y=0.1$  reported by Kuramaji and coworkers \cite{Kurumaji2015}. The  intermediate magnetization step at $H_{c*}$ for temperatures $T>20$~K may originate from a more delicate competition between the exchange couplings and thermal fluctuations at this particular Zn concentration $y=0.07$. On account of the XRD results on crushed single crystals, we discard the possibility that the intermediate steps are due to impurity phases.
While for $y=0.05$ no remanent magnetization has been reported down to 13~K and the antiferromagnetic ordering appeared close to 60~K, for $y=0.1$ the metastable state with a finite remanent magnetization appears also in the range 40-30~K and at a similar ordering temperature as in our case.

\section{Experimental Results}

\subsection
    {Temperature dependent polarized absorption spectra}

\begin{figure*}[t]
	\centering
	\includegraphics[width=\textwidth]{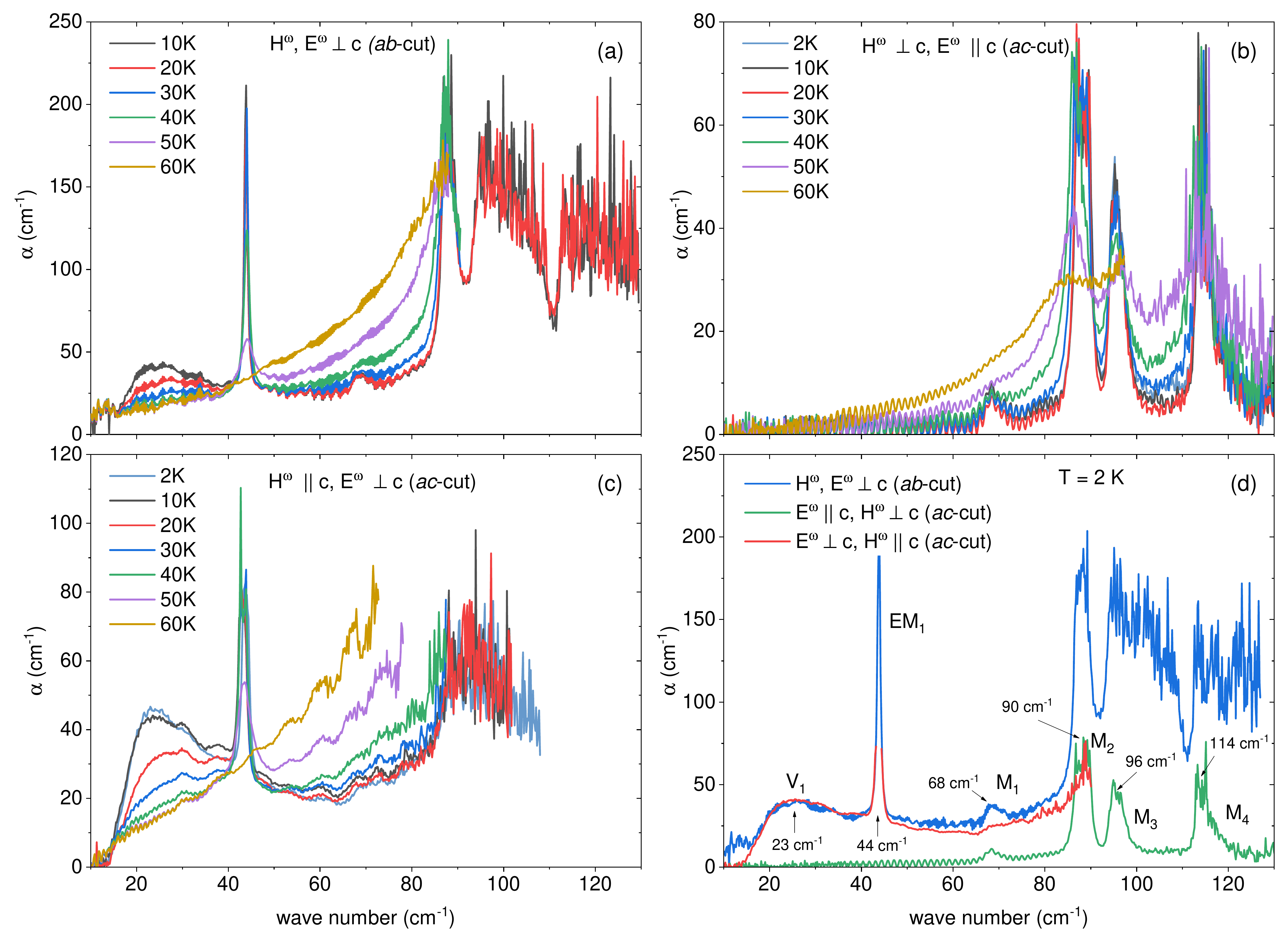}
	\caption{\label{fig:tempdep} Absorption spectra for several
	temperatures above and below $T_N$ for different polarization
	configurations: (a) $E^{\omega}\perp c, H^{\omega} \perp c$
	($ab$-cut sample) , (b)  $E^{\omega}\parallel c, H^{\omega} \perp c$
	($ac$-cut sample), (c)  $E^{\omega}\perp c, H^{\omega} \parallel c$
	($ac$-cut sample), and (d) comparison of the  absorption spectra at
	2\,K for the three different configurations.}
\end{figure*}

\indent The temperature dependence of the THz absorption
spectra for the polarization configuration
$E^{\omega}\perp c,H^{\omega}\perp c$ ($ab$-cut) and the
configurations $E^{\omega}\perp c, H^{\omega}\parallel c$
and $E^{\omega}\parallel c, H^{\omega}\perp c$ ($ac$-cut)
are shown in Fig.~\ref{fig:tempdep}(a)-(c), respectively. Above
$T_N$, there are no
detectable excitations. The strong monotonous increase in absorption
with increasing frequency for $E^{\omega}\perp c$ can be
attributed to the low-energy tail of the lowest-lying infrared-active
phonon of $E_1$-symmetry with an eigenfrequency of
130\,cm$^{-1}$ \cite{Reschke2020,Stanislavchuk2020}.  The decreasing absorption
of this contribution with decreasing temperature is in agreement with
the strong narrowing of the IR-active phonon, when the system is
cooled below $T_N$ \cite{Reschke2020}. Note that the lowest lying
IR-active phonon for $E^{\omega}\parallel c$ is located at around
200\,cm$^{-1}$ \cite{Reschke2020,Stanislavchuk2020} and, therefore, the frequency
range where transmission can be detected is wider than for
$E^{\omega}\perp c$ polarization. Upon cooling below
$T_N$, several new modes emerge or become observable
due to the reduced absorption of the IR-phonon contribution in the
investigated frequency range. Since all of
these modes seem to gain intensity and sharpen upon cooling we
assign them to new excitations of the magnetically ordered state.

\begin{table}[htb]
\squeezetable
\begin{ruledtabular}
\centering\footnotesize
   \begin{tabular}{cccccc}
    \multirow{3}{*}{\textbf{mode}}
	& {\bfseries $\boldsymbol{\omega_0}$} &
	\textbf{$\boldsymbol{ab}$-cut} &
	\multicolumn{2}{c}{\textbf{$\boldsymbol{ac}$-cut}} &
	\multirow{3}{*}{\textbf{activity}} \\
    & $[\mathrm{cm}^{-1}]$ & $E^{\omega}\perp c$ & $E^{\omega}\perp c$
    & \multicolumn{1}{c}
     {$E^{\omega}\parallel c$}  &  \\
    & &$H^{\omega}\perp c$ & $H^{\omega}\parallel c$
    & \multicolumn{1}{c}
     {$H^{\omega}\perp c$} &  \\
    \midrule
	 $V_1$ & 23 &  $\checkmark$ & $\checkmark$  &  $\times$
	 &  $E^{\omega}\perp c$     \\
	 $EM_1$ & 44 &  $\checkmark$  & $\checkmark$  &  $\times$
	 &  $E^{\omega}\perp c$     \\
	 $M_1$ & 68 &  $\checkmark$  & $\times$      &  $\checkmark$
	 &  $H^{\omega}\perp c$     \\
	 $M_2$ & 90 &  $\checkmark$  & n.r.    &  $\checkmark$
	 &   $H^{\omega}\perp c$     \\
	 $M_3$ & 96 & n.r. & n.r.    &  $\checkmark$
	 &  ?  	\\
	 $M_4$ & 114 & n.r. & n.r.    &  $\checkmark$
	 &  ?   \\\midrule
	 $EM$ & 40 & $\checkmark$ & $\checkmark$    &  $\times$
	 &  $E^{\omega}\perp c$ \\
	 $MM_1$ & 90 &  $\checkmark$  & $\times$    &  $\checkmark$
	 &  $H^{\omega}\perp c$     \\
    \end{tabular}
    \caption{\label{tab:selectionrules} Upper part: Selection rules for the
    excitations found in \fzmo{} for the different polarization configurations
    as shown in Fig. \ref{fig:tempdep}. The  notation $\checkmark$
    and $\times$ indicates the presence or absence of a mode.  The cases when
    no observation was possible are denoted as n.r. (not resolved). Lower
    part: reported excitations in the AFM phase of pure
    \fmo{} at 4.5~K taken from Ref.~\cite{Kurumaji2017a}.}
\end{ruledtabular}
\end{table}

Note that the high-frequency part of some spectra is not shown, because
the transmission values were too low to give reliable information in
this region. For the same reason the absorption maxima of the most
intense modes could not be resolved. Whenever this was the case, the
eigenfrequency $\omega_0$ was obtained by assuming a symmetric lineshape and
choosing the center of the absorption. For the broad band $V_1$ the frequency
of the intensity maximum is given as a characteristic frequency. \\
\indent A comparison of the spectra measured at 2~K for the three polarization
configurations is shown in Fig.~\ref{fig:tempdep}(d). The resulting selection
rules and eigenfrequencies for the detected modes are summarized in Table
\ref{tab:selectionrules} together with the modes reported for the AFM phase
of pure \fmo{} \cite{Kurumaji2017a}. The lowest-lying
broad asymmetric excitation band V$_1$ and mode $EM_1$ are observed
for the $ab$-cut as well as in the $ac$-cut spectrum with
$E^{\omega}\perp c,~ H^{\omega}\parallel c$ and are, therefore,
assigned to be electric-dipole active for $E^{\omega}\perp c$. The fact
that the intensity of $V_1$ does not alter for these two configurations
implies that it is not magnetic-dipole active. No feature similar to
the $V_1$-band has been reported for any other compound of the
\fzmox{} series. For $EM_1$ our data does not allow to exclude a
difference in intensity for the two configurations, because the maximum
of $EM_1$ was not resolved. However, the eigenfrequency and the selection rule
for $EM_1$ are in good agreement with a reported excitation called $EM$ in
pure \fmo{} \cite{Kurumaji2017a} (see Table~\ref{tab:selectionrules}),
where the intensities for these two configurations
seem to show no differences. Therefore, we also discard a possible
magnetic-dipole activity of $EM_1$. In addition,  $EM_1$ was not observed
for a Zn concentration of $y=0.125$, where the AFM phase is completely
suppressed and the ground state is ferrimagnetic \cite{Kurumaji2017a}. \\
\begin{figure*}[t]
    \centering
    \includegraphics[width=\linewidth]
    {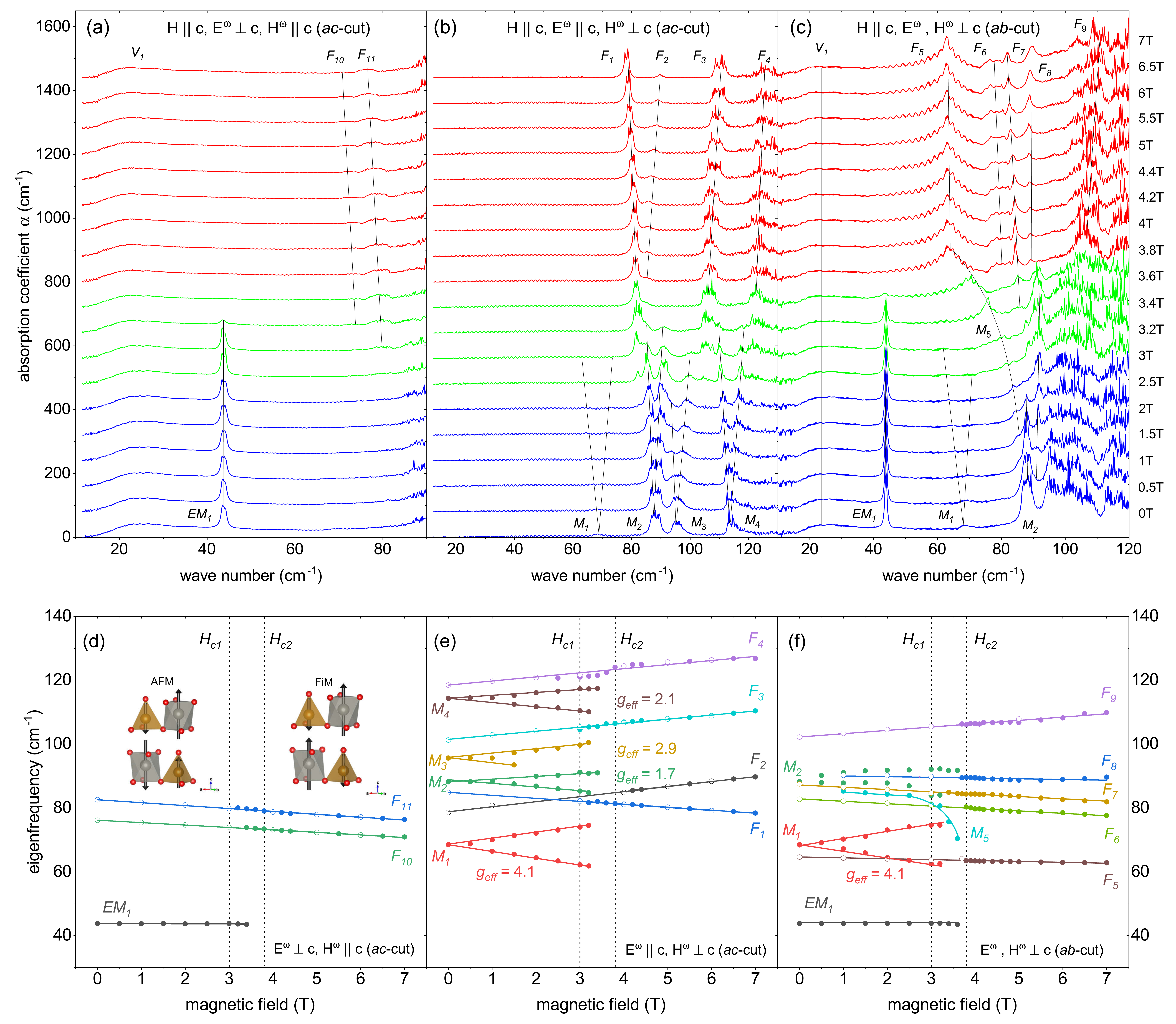}
    \caption{\label{fig:fielddep} Upper panels: Magnetic field
    dependence of the absorption spectra  with $H\parallel c$ at
    $T=13$\,K in the range of 0\,T to 7\,T for (a)
    $E^{\omega}\perp c, H^{\omega} \parallel c$ ($ac$-cut),
    (b) $E^{\omega}\parallel c, H^{\omega}
    \perp c$ ($ac$-cut) and (c) $E^{\omega},H^{\omega}\perp c$
    ($ab$-cut). All spectra are shown with a
    constant offset of 80\,cm$^{-1}$ with respect to the 0~T
    spectra. Spectra in the AFM phase are plotted in blue, for
    $H_{c1}<H<H_{c2}$ in green, and in the ferrimagnetic phase
    in red. The solid lines are drawn to guide the eyes. Lower
    panels: Magnetic field dependence of the eigenfrequencies for
	all detected modes (except A$_1$) for (d)
	$E^{\omega}\perp c, H^{\omega} \parallel c$ ($ac$-cut),
	(e) $E^{\omega}\parallel c, H^{\omega} \perp c$ ($ac$-cut)
	and (f) $E^{\omega},H^{\omega}\perp c$ ($ab$-cut).
	Filled symbols represent data obtained in increasing field
	starting from 0~T, while open symbols were taken upon decreasing
	fields. Dotted lines represent the critical fields of the
	magnetic transition taken from the magnetization data The inset in (d)
	depicts the suggested collinear spin configurations of the AFM and
	FiM states.}
\end{figure*}
\indent In contrast to modes $V_1$ and $EM_1$, the four additional modes,
$M_1$-$M_4$, observed at higher frequencies can be excited
for $H^{\omega}\perp c$ and are thus considered to be magnetic dipole
active. An additional electric-dipole activity is discarded for
excitation $M_1$, because it does not appear for the configuration
$E^{\omega}\perp c,~ H^{\omega}\parallel c$.
For modes $M_2$-$M_4$ our spectra for
$E^{\omega}\perp c,~ H^{\omega}\parallel c$ do not allow us
to unambiguously exclude an electric-dipole activity, but we
expect any such contribution to be weak. Note, that a
magnetic-dipole active mode (named $MM_1$, see see
Table~\ref{tab:selectionrules}) with an eigenfrequency close to
$M_2$ has been reported for $x=0$ and $x=0.125$, previously
\cite{Kurumaji2017a} and determined to be only magnetic-dipole active.
Modes $M_1$, $M_3$, and $M_4$ have not been observed before, probably
due to the low intensity of $M_1$ and the fact that the frequency
range of $M_3$ and $M_4$ was not resolved in previous studies.

\subsection{Magnetic field dependent absorption spectra}

\begin{figure}
    \centering
    \includegraphics[width=\linewidth]{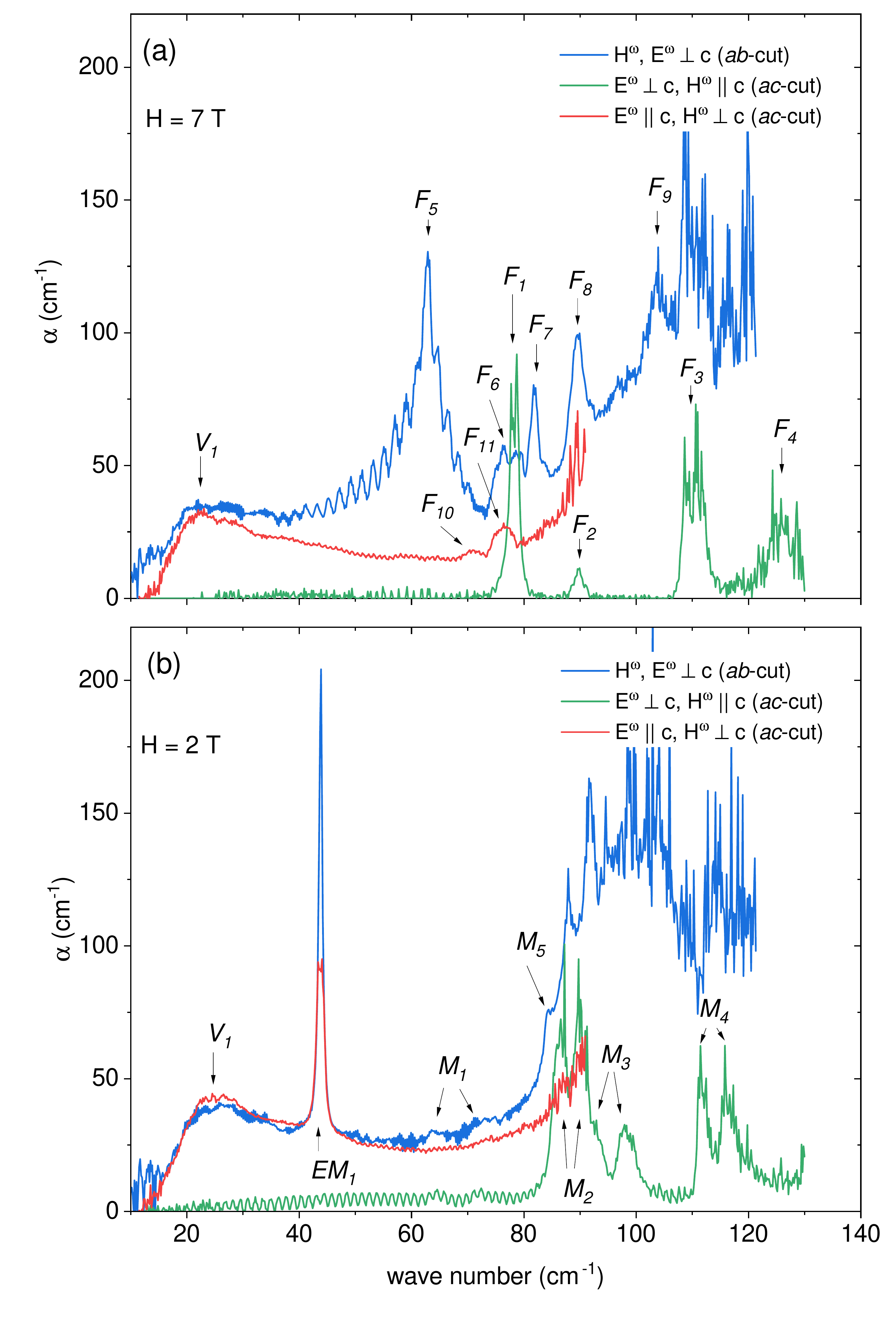}
    \caption{ \label{fig:CompFM_AFM} Comparison of the absorption
    spectra (a) at $H=7\,$T in the FiM state and (b) at $H=2\,$T in the
    AFM state for the three polarization configurations. The
    splitting of the AFM modes $M_{1}-M_{4}$ is indicated in (b).}
\end{figure}

The magnetic field dependence of the absorption spectra for
the three possible polarization configurations measured for $H\parallel c$
\textcolor{blue}{is} shown in Fig.~\ref{fig:fielddep}(a)-(c). Since the transition
to the ferrimagnetic state occurs when the magnetic field is parallel to the
$c$-axis, the $ab$-cut sample was measured in Faraday configuration
($H\parallel c,\; k\parallel c$) and the $ac$-cut in Voigt
configuration ($H\parallel c,\; k\perp c$), where $k$ is the wave vector of
the incoming THz beam. Spectra measured with $H\perp c$ (not shown here)
exhibited no field dependence of the absorption spectra. \\
\indent The measurements were performed at 13\,K, where the critical fields
$H_{c1}$ and $H_{c2}$ of the magnetization steps and the saturation
magnetization (see Fig.~\ref{fig:magnetization}(b)) of the compound
were accessible using our experimental setup. When
increasing the magnetic field in the antiferromagnetic phase (blue spectra)
we observe a linear splitting of the modes $M_1-M_4$ as indicated by the
black lines in Fig.~\ref{fig:fielddep}(b) in Voigt configuration and for
$M_1$ also in Faraday configuration (Fig.~\ref{fig:fielddep}(c)). The slope of the linear dependence was parametrized by an  effective $g$-factor defined by $\hbar\omega=\hbar\omega_0\pm g_{\mathrm{eff}}\mu_B H$
for both branches. The values are indicated in
Fig.~\ref{fig:fielddep}(e) and (f). The behaviour of $M_2$ in Faraday
configuration is more complex, as a low-frequency satellite $M_5$ seems
to emerge and soften with increasing field up to $H_{c2}$. For the
electric-dipole active mode $EM_1$ no shift or splitting with increasing
magnetic field could be resolved, but it looses
intensity when approaching the transition region to the FiM state  and it
finally disappears for $H>H_{c2}$ (spectra for $H_{c1}<H<H_{c2}$ are shown
in green, for $H>H_{c2}$ in red). This behaviour is in agreement
with the corresponding $EM$-mode in pure \fmo{} \cite{Kurumaji2017a}.
Above $H_{c2}$ modes $M_1-M_4$ also
vanish, and new modes appear in the ferrimagnetic phase. We want to
emphasize that the coexistence of modes of the AFM and FiM phases for
$H_{c1}<H<H_{c2}$ indicates a two-phase regime in this field region.
These new modes also show a field dependence and are labeled $F_1-F_{11}$.
A detailed comparison of the splitting of the antiferromagnetic modes
at 2~T and the shape of the ferrimagnetic
modes at 7~T is given in Fig.~\ref{fig:CompFM_AFM}.
Notably, the broad band $V_1$ does not exhibit any significant changes
with increasing magnetic fields and does not seem to be influenced by the
transition to the FiM state.

The resonance frequencies of the ferrimagnetic modes
$F_1-F_{11}$ either increase or decrease linearly with increasing
magnetic fields as shown in Fig.~\ref{fig:fielddep}. Additionally,
we included the resonance frequencies
of modes $F_1-F_{11}$ measured upon lowering the magnetic field to zero
(spectra not shown here) as
open symbols in Fig.~\ref{fig:fielddep}(d)-(f). The eigenfrequencies of
$F_1$ and $F_2$ were found to cross each other below $H_{c1}$. The
eigenfrequencies of $F_1-F_{11}$ in  the zero-field FiM state are given
in Table~\ref{tab:FMRselectionrules}, together with the effective $g$-factors.
In the lower part of Table~\ref{tab:FMRselectionrules} we
list all reported modes of the FiM phase of \fzmox{} for a comparison.
For $y=0$ and $y=0.125$ a magnetic-dipole active mode
(called $\nu_3$ or $MM_2$) with a similar eigenfrequency as $F_1$ has been
reported \cite{Kurumaji2017a}. For $y=0.25$ and  $y=0.4$ this
mode $\nu_3$ seems to have developed an additional electric-dipole activity
\cite{Kurumaji2017}. The additional modes $\nu_1$, $\nu_2$, and $\nu_4$ at lower
frequencies cannot be directly related to the modes observed in our study,
as we expect a continuous evolution of the modes with increasing Zn
content and, therefore, consider the reported spectra for $y=0$ and
$y=0.125$ as the most relevant.

\section{Discussion}

\subsection{Modes \texorpdfstring{$EM_1$}{EM1} and
\texorpdfstring{$M_1-M_5$}{M1-M4} of the AFM phase}

\indent As already mentioned above, modes $EM_1$ and $M_2$ have been
reported already for pure   \fmo{} and interpreted as precessional
modes of the collinear AFM structure, both being doubly degenerate
modes \cite{Kurumaji2017a}.
While the degeneracy of $M_2$ is clearly
lifted in the external magnetic field with $g_{\text{eff}}=1.7$ for
$H^{\omega}\parallel a$, the possible doublet nature of $EM_1$ remains
to be confirmed in higher magnetic fields. The respective electric and
magnetic dipole activity of the two modes was assigned to an inverse
Dzyaloshinskii-Moriya mechanism and different single-ion anisotropies
of tetrahedral and octahedral sites \cite{Kurumaji2017a}. \\
\indent As our spectra reveal three additional magnetic modes
$M_1, M_3, M_4$, we have to take into account that the disorder
induced by substitution of iron by zinc may result in an increased number
of nonequivalent tetrahedral Fe sites. In early
M\"{o}ssbauer studies for
various zinc dopings at least four different tetrahedral iron sites with
different hyperfine fields could be distinguished \cite{Varret1972}.
Given the absence of $M_1$ in the spectra of pure \fmo{}
\cite{Kurumaji2017a} and the low intensity of the mode, we assume
that this mode is due to the Zn induced disorder.
As the spectral range reported for \fmo{} was
limited to about 94~cm$^{-1}$ \cite{Kurumaji2017a}, it is not clear,
whether $M_3$ (at 96~cm$^{-1}$) and $M_4$ (at 114~cm$^{-1}$) are
also disorder-induced modes or whether they are present in pure \fmo{},
too.
Preliminary THz measurements on pure \fmo{}
\cite{Strinic2020}, however, indicate that an excitation with a similar eigenenergy as $M_4$ is present also in pure \fmo{} and may be described as an intrinsic
mode of \fmo{}. Under these assumptions, we conclude that at least the
three modes $EM_1,M_2,M_4$ are to be regarded inherent to the magnetic
structure of the pure compound and the two doubly degenerate modes
$M_1,M_3$ may emerge due to the dilution of the magnetic iron sites
by non-magnetic zinc. Since mode $M_5$ shifts to lower eigenfrequencies
with increasing magnetic field and disappears above $H_{c2}$, its
behavior may be interpreted as a soft mode of the magnetic phase
transition. However, its relation to $M_2$, from where it seems to
originate, remains unresolved at present.
Besides the influence of different Fe sites, the large number of
magnetic modes may also be due to additional spin stretching modes
\cite{Penc2012, Kocsis2018a}.

\begin{table*}
\squeezetable
\begin{ruledtabular}
    \centering\footnotesize
    \begin{tabular}{lcccccc}
    \multirow{2}{*}{\textbf{mode}}
    & {\bfseries $\boldsymbol{\omega_0}$
       ($\boldsymbol{H}^{\boldsymbol{dc}}
       \boldsymbol{=0}$\,\textbf{T})}
    & $g_{\mathrm{eff}}$
	& {\textbf{$\boldsymbol{ab}$-cut}} &
	\multicolumn{2}{c}{\textbf{$\boldsymbol{ac}$-cut}} &
	\multirow{2}{*}{\textbf{activity}} \\
	
    &  $[\mathrm{cm}^{-1}]$ &  &
    $E^{\omega},H^{\omega}\perp c$
    & $E^{\omega}\perp c, H^{\omega}\parallel c$
    & \multicolumn{1}{c}
     {$E^{\omega}\parallel c, H^{\omega}\perp c$} &  \\

    \midrule

     $F_1$    & 85  & -2.0  &  $\times$     & $\times$
	    & $\checkmark$      & ?
	    \\
	 $F_2$    & 79  &  3.4  &  $\times$     & $\times$
	    & $\checkmark$      & ?
	    \\
	 $F_3$    & 102 &  2.7  &  $\times$    &  n.r.
	    & $\checkmark$      & ?
	    \\
	 $F_4$    & 119 &  2.7  &  n.r.              & n.r.
	    & $\checkmark$      &  ?
	    \\
	 $F_5$    & 65  & -0.6  & $\checkmark$       & $\times$
	    & $\times$          & ?
	    \\
	 $F_6$    & 83  & -1.6  & $\checkmark$       & $\times$
	    & $\times$          & ?
	    \\
	 $F_7$    & 87  & -1.7  & $\checkmark$       & $\times$
	    & $\times$          & ?
	    \\
	 $F_8$    & 91  & -0.2  & $\checkmark$       & n.r.
	    & $\times$          & ?
	    \\
	 $F_9$    & 102 &  2.3  & $\checkmark$       & n.r.
	    & $\times$          & ?
	    \\
	 $F_{10}$ & 76  & -1.7  &  $\times$              &  $\checkmark$
	    & $\times$          &  ?
	    \\
	 $F_{11}$ & 83  & -1.9  & $\times$&  $\checkmark$
	    & $\times$          &  ?  	
	    \\
	 \midrule
	 $MM2$ $(y=0)$            & 87 & -2.0 & - & $\times$     &
	 $\checkmark$ & $H^{\omega}\perp c$   	
	    \\
	 $MM2, \nu_3$ $(y=0.125)$ & 87 & -2.5 & - & $\times$     &
	 $\checkmark$ & $H^{\omega}\perp c$   	
	    \\
	 $\nu_1$ $(y=0.25)$       & 47 &  4.0 & - & $\times$     &
	 $\checkmark$ & $H^{\omega}\perp c$  	
	    \\
	 $\nu_2$ $(y=0.25)$       & 76 &  2.6 & - & $\times$     &
	 $\checkmark$ & $H^{\omega}\perp c$  	
	    \\
	 $\nu_3$ $(y=0.25)$       & 87 & -2.6 & - & $\checkmark$ &
	 $\checkmark$ & $E^{\omega},H^{\omega}\perp c$   	
	    \\
	 $\nu_1$ $(y=0.4)$        & 47 &  4.0 & - & $\checkmark$ &
	 $\checkmark$ & $E^{\omega},H^{\omega}\perp c$  	
	    \\
	 $\nu_2$ $(y=0.4)$        & 73 &  3.6 & - & $\times$     &
	 $\checkmark$ & $H^{\omega}\perp c$  	
	    \\
	 $\nu_3$ $(y=0.4)$        & 87 & -3.1 & - & $\checkmark$ &
	 $\checkmark$ & $E^{\omega},H^{\omega}\perp c$   	
	    \\
	 $\nu_4$ $(y=0.5)$        & 42 &  -   & - & $\times$ &
	 $\checkmark$ & $H^{\omega}\perp c$   	
	    \\
    \end{tabular}
    \caption{\label{tab:FMRselectionrules} Selection rules for the
    excitations $F_1-F_{11}$  found for the different polarization
    configurations at 13~K as shown in Fig. \ref{fig:tempdep}. The
    eigenfrequencies $\omega_0$ correspond to the zero-field values in
    the FiM phase. The sign of  $g_{\mathrm{eff}}$ indicates the sign
    of the linear slope with increasing magnetic field. The  notation
    $\checkmark$ and $\times$ indicates the presence or absence of a
    mode. The cases when no observation was possible are
    denoted as n.r. (not resolved). The lower part of the table lists
    reported modes of the FiM phase for different Zn concentrations and their assigned optical activity
    taken from Refs. \cite{Kurumaji2017a, Kurumaji2017, Yu2018}. }
\end{ruledtabular}
\end{table*}

\subsection{Modes of the ferrimagnetic phase}

\indent At present we identify eleven modes $F_1-F_{11}$ in the
spectra of the FiM phase for all measured polarization configuration
above $H_{c2}$ (see Tab.~\ref{tab:FMRselectionrules}). In principle,
this corresponds to the number of excitation branches in the AFM phase
below $H_{c1}$, if the $EM_1$ mode is assumed to be doubly degenerate
and $M_5$ corresponds to a single mode.

A direct comparison of the eigenfrequencies in zero-field and
the effective $g$-factors estimated by $\hbar\omega=\hbar\omega_0 +  g_{\mathrm{eff}}\mu_BH$
is provided in Table~\ref{tab:FMRselectionrules},
together with the FiM modes previously reported for \fzmox{}
\cite{Kurumaji2017a, Kurumaji2017, Yu2018}. The corresponding effective
$g$-factors of these modes were determined by us using the published data.

To estimate the influence of demagnetizing fields for the investigated samples we model the samples as ellipsoids
\cite{Osborn1945}. The $ab$-cut sample is assumed to have semi-axes $a,b,c$ corresponding to its approximate dimensions $2\times 1.5\times0.6$~cm$^3$ and experiences a demagnetization factor $N_{z}=0.623$ along the $c$-axis. The $ac$-cut sample with  $2\times 1.6\times 1.48$~cm$^3$ yields $N_{z}=0.343$. The resulting demagnetization fields $\mu_0N_zM_S$, where $M_S\simeq 0.85 \mu_\mathrm{B}/\mathrm{f.u.}$ denotes the magnetization value measured at 13~K in a field of 5~T as shown in Fig.~\ref{fig:magnetization}(b), correspond then to 0.05~T and 0.03~T for the $ab$-cut and $ac$-cut samples, respectively. Since the resulting fields are about two orders of magnitude smaller than the applied external fields in the FiM state, we neglect demagnetization effects in the following.

Unfortunately, it is not possible to determine clear selection rules for the
FiM modes, either because the excitations could not be resolved experimentally
or their presence or absence in the three polarization configuration does not
yield a consistent selection rule. For example, mode $F_1$ agrees well both
in eigenfrequency and $g$-factor to the magnetic-dipole active mode $MM_2$
reported for $y=0$ and $y=0.125$ \cite{Kurumaji2017a}, where the selection
rule $H^{\omega}\perp c$ was established by investigating an $ac$-cut sample
in Voigt configuration, only. From our measurement in Voigt configuration alone,
we would derive the same selection rule, but the absence of this mode in our
$ab$-cut measurements with $H^{\omega}\perp c$ does not allow to establish this
selection rule. Similarly, modes $F_5-F_9$ should appear in one of the
configurations in the $ac$-cut measurements to establish a selection rule,
but there seem to be no corresponding excitations.

In principle, the symmetry of the FiM phase was found to allow the occurrence
of gyrotropic birefringence as reported for excitations $\nu_2$ and $\nu_3$
for Zn concentrations of $y=0.25$ and $y=0.4$ \cite{Kurumaji2017}, which seem
to have no direct correspondence to the mode observed in this study. However,
we can not exclude that similar magnetoelectric effects might be present for
some of these modes.

\subsection
    { Modelling the absorption band \texorpdfstring{${V_1}$}{V1} in terms of a vibronic mode}

\begin{figure}[bt]
    \centering
    \includegraphics[width=\linewidth]
        {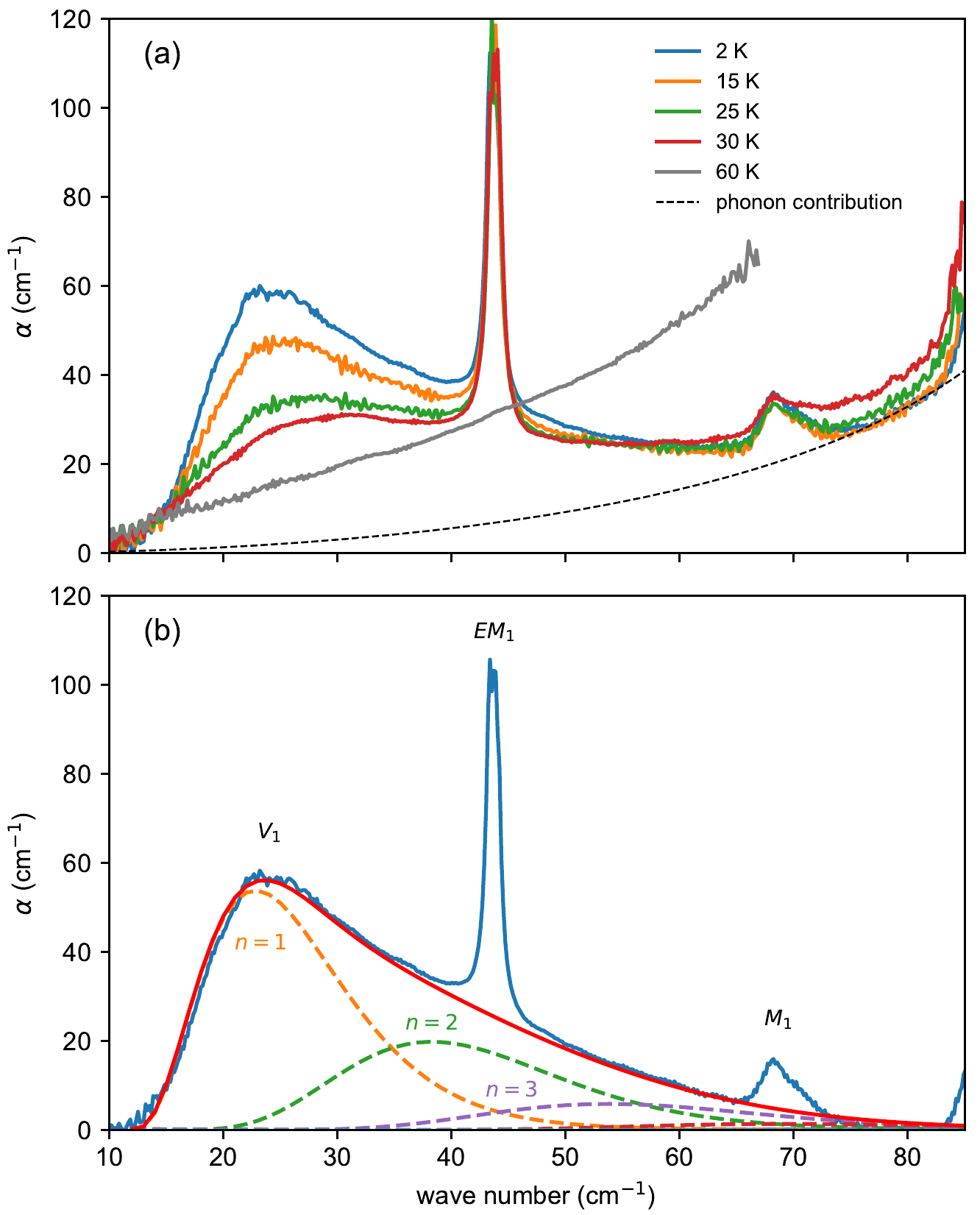}
    \caption{\label{fig:vibronic}(a) THz absorption data at various
    temperatures for $E^{\omega}\perp c,H^{\omega}\perp c$, and a
    Lorentzian phonon contribution of the lowest-lying phonon as
    described in the text (black dashed line), (b) THz absorption
    spectrum at 2\,K, where the phonon contribution was subtracted,
    and a simulation using Eqs.  (\ref{eq:sum}) and  (\ref{eq:coupling}),
    including the contributions for $n=1,2,3$ (dashed lines).}
\end{figure}

\indent In the following we will analyze the nature of the
electric-dipole active excitation band $V_1$, which emerges and
gains intensity with decreasing temperatures below $T_N$ (see
Fig.~\ref{fig:vibronic}(b)). It can only be observed for
$E^{\omega}\parallel a$ and does not exhibit any changes
in magnetic fields up to 7~T. The onset of the
excitation occurs at about 10\,cm$^{-1}$,
which is in agreement with the reported energy difference between
the ground and first excited state of the Fe$^{2+}$ ions in
tetrahedral environment for pure \fmo{}  \cite{Varret1972}.

Strong absorption features related to transitions within the low-lying electronic $d-d$-levels of Fe$^{2+}$ ions in
tetrahedral sites have been reported for diluted Fe$^{2+}$ on tetrahedral sites in semiconductors \cite{Testelin1992} and also in several compounds with concentrated Fe$^{2+}$ ions, e.g. in the spinels FeSc$_2$S$_4$ \cite{Mittelstaedt2015,Laurita2015a} and  FeCr$_2$S$_4$ \cite{Strinic2020a} or in the system Sr$_2$FeSi$_2$O$_7$ \cite{Mai2016}. Moreover, electron-phonon coupling can lead to vibronic excitations, where electronic and
vibrational degrees of freedom cannot be separated. In particular, the fact that the electronic eigenfrequencies of the tetrahedrally coordinated Fe$^{2+}$ ions are of the same order of magnitude as possible involved phonon modes may foster vibronic effects
\cite{Vallin1970, Vogel1980, Wittekoek1973, Feiner1982}. A corresponding analysis of such low-lying vibronic excitations has been elaborated, e.g., by Testelin and coworkers \cite{Testelin1992}.

Here, we follow this line and concentrate on modelling the lineshape of the absorption by using the general approach to describe  vibronic
excitations in case of a linear electron-phonon coupling
\cite{Stoneham1975, Toyozawa2003}. The normalized absorption spectrum
at $T=0$ can be described by
\begin{equation}
    I(\omega)=\int\limits_{-\infty}^{\infty}f(t)
    e^{i\omega t} \mathrm{d}t
\end{equation}
with the generating function
\begin{equation}
    f(t)=\exp(-i\omega_0t - S +S(t))
\end{equation}
where
\begin{equation}
    S(t)=\int\limits_{0}^{\infty}s(\omega) e^{-i\omega t}
    \mathrm{d}\omega
\end{equation}
is determined by the coupling function $s(\omega)$,
which describes the distribution of phonon modes in a
given material and $S=S(t=0)$.
Expanding $f(t)$ in a power series in $S(t)$ results in
\begin{equation}\label{eq:sum}
    I(\omega)=e^{-S}\sum\limits_{n=0}^{\infty}\frac{S^n }{n!}
    \Gamma_n(\omega)
\end{equation}
with
\begin{equation}
    \Gamma_n(\omega)=\int\limits_{-\infty}^{\infty}
    e^{i(\omega-\omega_0)t}\left[\frac{S(t)}{S}\right]^n\mathrm{d}t
\end{equation}
which can be calculated iteratively for $n>1$ by the convolution
\begin{equation}
    \Gamma_n(\omega)=\int\limits_{-\infty}^{\infty}
    \Gamma_{n-1}(\omega^\prime)\Gamma_{1}(\omega +\omega_0
    -\omega^\prime)\mathrm{d}\omega^\prime
\end{equation}
\noindent The number $n$ corresponds to the number of phonons involved
in the absorption process \cite{Stoneham1975} and, hence, the
zero-phonon-line with $n=0$ describes the purely electronic
transition at $\omega_0$, which is not visible in our spectra.

In order to simulate the vibrational lineshape the choice of a suitable coupling function is important. Optically active phonons are usually at higher
frequencies than the observed absorption band and, therefore, we consider a coupling to acoustic phonons, which are usually modelled by a Debye-like phonon density of states. Using an exponential decay  around the Debye frequency
$\omega_1$ instead of a sharp cut-off \cite{Pullerits1995} leads to the first term in our coupling function

\begin{equation}\label{eq:coupling}
    s(\omega)=\exp\left(-\frac{\omega}{\omega_1}\right)
    \frac{\omega^2S_1}{2\omega_1^3}+
    \exp\left(-\frac{\omega}{\omega_2}\right)
    \frac{\omega^3S_2}{6\omega_2^4}.
\end{equation}
This first Debye-like term alone, however, does not allow to fully capture the entire absorption band. Therefore, the second term with a cut-off frequency $\omega_2$ is introduced as an empirical modification of the phonon-density of states, satisfying
the condition $S=S_1+S_2$ as required \cite{Toyozawa2003}. Note that this second term was suggested to model molecular vibronic spectra with the same approach used here \cite{Ratsep2014}. The choice of the second term is somewhat arbitrary and other terms satisfying $S=S_1+S_2$ might give the same result. As the experimental phonon-density of states is not available for comparison with the coupling function, this term and its additional two parameters deem, however, necessary to describe the entire absorption curve successfully as we discuss in the following.

In order to analyze the lineshape of the $V_1$ excitation, we use a
spectrum obtained for $E^{\omega}\perp c,H^{\omega}\perp c$
($ab$-cut sample)  and subtract the high-frequency
contribution stemming from the lowest-lying infrared-active phonon
modelled by a Lorentzian lineshape with eigenfrequency
$\omega_0=128.6$\,cm$^{-1}$, damping $\gamma=9.0$\,cm$^{-1}$, and ionic
plasma frequency $\omega_p=158.4$\,cm$^{-1}$ (see
Fig.~\ref{fig:vibronic}(a)) in agreement with reported data  for pure
\fmo{} \cite{Reschke2020}.
As the narrow electric-dipole excitation
$EM_1$ on top of the $V_1$-band is suppressed in the ferrimagnetic
phase above $H_{c2}$ (see Fig.~\ref{fig:fielddep}) and $M_1$ is
clearly magnetic in origin, while $V_1$ does not alter, we do
not consider any coupling of these excitations. \\
\indent To restrict the parameters for this simulation, we fixed the
zero-phonon-line to $\omega_0=12$\,cm$^{-1}$, a value close to
experimental  \cite{Reschke2020,Stanislavchuk2020} and theoretical
estimates \cite{Varret1972}
and minimized the number of phonons involved in the process to $n=3$.
Using these constraints, we could reproduce the experimental lineshape
with the five parameters $S_1=0.02$, $S_2=1.1$, $\omega_1=12$\,cm$^{-1}$,
$\omega_2=3.6$\,cm$^{-1}$ and the amplitude $A=2.4\cdot 10^{3}$\,cm$^{-1}$, which is
defined by
$\alpha(\omega)=A\cdot I(\omega)$. The agreement between data and
simulation is very good and, hence, we conclude that this simple approach
satisfactorily reproduces the $V_1$ band and supports a vibronic
origin of the band. Note, that higher convolutions for $n>3$ do only
weakly modify the high-frequency tail. The values for $\omega_1$ and $\omega_2$ must be regarded
as a parametrization of the acoustic phonon branches coupled to the
electronic transitions. As such a vibronic band has not been
reported for pure \fmo{} \cite{Kurumaji2017a}, we assume that the
disorder introduced by the Zn ions and the corresponding impurity
modes are responsible for the occurrence of this absorption band.
It remains to be clarified why THz studies of comparable samples with $y>0$
do not report a similar feature
\cite{Kurumaji2017a,Kurumaji2017,Yu2018} and why this mode can only be observed below $T_N$.

A possible explanation for the emergence below $T_N$ could
be that long range magnetic ordering triggers the splitting of the lowest-lying
iron orbitals as suggested in Ref.~\cite{Varret1972}. Moreover, some of us recently reported the emergence of electronic excitations in the far- and mid-infrared regime related to the higher-lying $d-d$-transition of the Fe$^{2+}$ ions at tetrahedral sites in pure \fmo{} below $T_N$, where energy scales of 13~cm$^{-1}$ and 26~cm$^{-1}$ occur as a difference between the observed transitions. This corroborates our interpretation that the vibronic band involves the lowest-lying electronic states of Fe$^{2+}$ ions at tetrahedral sites.

\section{Summary}
We observed ten magnetic modes in the AFM state in \fzmo{} in finite magnetic fields,
including both magnetic- and electric-dipole active modes. Assuming that the
lowest-lying mode is a doublet with an unresolved splitting in the magnetic fields
applied in this study, the actual number of AFM modes would increase to eleven. This
is the number of modes observed in the magnetic field induced FiM state in \fzmo{}.
The large number of modes, far exceeding the number of magnetic
sublattices in pure \fmo{}, may imply that the number of magnetic sublattices
is increased due to Zn substitution on the tetrahedral sites. Another possible
reason for the large number of excitations is that spin-stretching modes also
become optically active, besides the
precessional modes described by linear spin-wave theory. In the transition
region between the two magnetic phases a coexistence of AFM and FiM modes is
present, implying  a coexistence of AFM and FiM phases for $H_{c1}<H<H_{c2}$.
Additionally, a broad electric-dipole active excitation band was observed in the
AFM phase and our analysis strongly suggests that it is of vibronic origin involving the lowest-lying
electronic $d$-states of Fe in tetrahedral environment, which are split by about
12~cm$^{-1}$.

\begin{acknowledgments}
We acknowledge support by the Deutsche Forschungsgemeinschaft via
TRR 80 (project no. 107745057) and via the Institutional Project
20.80009.5007.19 (Moldova).
\end{acknowledgments}

\bibliographystyle{apsrev4-2}

\begin{thebibliography}{40}%
\makeatletter
\providecommand \@ifxundefined [1]{%
 \@ifx{#1\undefined}
}%
\providecommand \@ifnum [1]{%
 \ifnum #1\expandafter \@firstoftwo
 \else \expandafter \@secondoftwo
 \fi
}%
\providecommand \@ifx [1]{%
 \ifx #1\expandafter \@firstoftwo
 \else \expandafter \@secondoftwo
 \fi
}%
\providecommand \natexlab [1]{#1}%
\providecommand \enquote  [1]{``#1''}%
\providecommand \bibnamefont  [1]{#1}%
\providecommand \bibfnamefont [1]{#1}%
\providecommand \citenamefont [1]{#1}%
\providecommand \href@noop [0]{\@secondoftwo}%
\providecommand \href [0]{\begingroup \@sanitize@url \@href}%
\providecommand \@href[1]{\@@startlink{#1}\@@href}%
\providecommand \@@href[1]{\endgroup#1\@@endlink}%
\providecommand \@sanitize@url [0]{\catcode `\\12\catcode `\$12\catcode
  `\&12\catcode `\#12\catcode `\^12\catcode `\_12\catcode `\%12\relax}%
\providecommand \@@startlink[1]{}%
\providecommand \@@endlink[0]{}%
\providecommand \url  [0]{\begingroup\@sanitize@url \@url }%
\providecommand \@url [1]{\endgroup\@href {#1}{\urlprefix }}%
\providecommand \urlprefix  [0]{URL }%
\providecommand \Eprint [0]{\href }%
\providecommand \doibase [0]{https://doi.org/}%
\providecommand \selectlanguage [0]{\@gobble}%
\providecommand \bibinfo  [0]{\@secondoftwo}%
\providecommand \bibfield  [0]{\@secondoftwo}%
\providecommand \translation [1]{[#1]}%
\providecommand \BibitemOpen [0]{}%
\providecommand \bibitemStop [0]{}%
\providecommand \bibitemNoStop [0]{.\EOS\space}%
\providecommand \EOS [0]{\spacefactor3000\relax}%
\providecommand \BibitemShut  [1]{\csname bibitem#1\endcsname}%
\let\auto@bib@innerbib\@empty
\bibitem [{\citenamefont {Spaldin}\ and\ \citenamefont
  {Ramesh}(2019)}]{Spaldin2019}%
  \BibitemOpen
  \bibfield  {author} {\bibinfo {author} {\bibfnamefont {N.~A.}\ \bibnamefont
  {Spaldin}}\ and\ \bibinfo {author} {\bibfnamefont {R.}~\bibnamefont
  {Ramesh}},\ }\href {https://doi.org/10.1038/s41563-018-0275-2} {\bibfield
  {journal} {\bibinfo  {journal} {Nature Materials}\ }\textbf {\bibinfo
  {volume} {18}},\ \bibinfo {pages} {203} (\bibinfo {year} {2019})}\BibitemShut
  {NoStop}%
\bibitem [{\citenamefont {Dong}\ \emph {et~al.}(2019)\citenamefont {Dong},
  \citenamefont {Xiang},\ and\ \citenamefont {Dagotto}}]{Dong2019}%
  \BibitemOpen
  \bibfield  {author} {\bibinfo {author} {\bibfnamefont {S.}~\bibnamefont
  {Dong}}, \bibinfo {author} {\bibfnamefont {H.}~\bibnamefont {Xiang}},\ and\
  \bibinfo {author} {\bibfnamefont {E.}~\bibnamefont {Dagotto}},\ }\href
  {https://doi.org/10.1093/nsr/nwz023} {\bibfield  {journal} {\bibinfo
  {journal} {National Science Review}\ }\textbf {\bibinfo {volume} {6}},\
  \bibinfo {pages} {629} (\bibinfo {year} {2019})}\BibitemShut {NoStop}%
\bibitem [{\citenamefont {K{\'{e}}zsm{\'{a}}rki}\ \emph
  {et~al.}(2015)\citenamefont {K{\'{e}}zsm{\'{a}}rki}, \citenamefont {Nagel},
  \citenamefont {Bord{\'{a}}cs}, \citenamefont {Fishman}, \citenamefont {Lee},
  \citenamefont {Yi}, \citenamefont {Cheong},\ and\ \citenamefont
  {R{\~{o}}{\~{o}}m}}]{Kezsmarki2015}%
  \BibitemOpen
  \bibfield  {author} {\bibinfo {author} {\bibfnamefont {I.}~\bibnamefont
  {K{\'{e}}zsm{\'{a}}rki}}, \bibinfo {author} {\bibfnamefont {U.}~\bibnamefont
  {Nagel}}, \bibinfo {author} {\bibfnamefont {S.}~\bibnamefont
  {Bord{\'{a}}cs}}, \bibinfo {author} {\bibfnamefont {R.~S.}\ \bibnamefont
  {Fishman}}, \bibinfo {author} {\bibfnamefont {J.~H.}\ \bibnamefont {Lee}},
  \bibinfo {author} {\bibfnamefont {H.~T.}\ \bibnamefont {Yi}}, \bibinfo
  {author} {\bibfnamefont {S.~W.}\ \bibnamefont {Cheong}},\ and\ \bibinfo
  {author} {\bibfnamefont {T.}~\bibnamefont {R{\~{o}}{\~{o}}m}},\ }\href
  {https://doi.org/10.1103/PhysRevLett.115.127203} {\bibfield  {journal}
  {\bibinfo  {journal} {Phys. Rev. Lett.}\ }\textbf {\bibinfo {volume} {115}},\
  \bibinfo {pages} {127203} (\bibinfo {year} {2015})}\BibitemShut {NoStop}%
\bibitem [{\citenamefont {Jungwirth}\ \emph {et~al.}(2018)\citenamefont
  {Jungwirth}, \citenamefont {Sinova}, \citenamefont {Manchon}, \citenamefont
  {Marti}, \citenamefont {Wunderlich},\ and\ \citenamefont
  {Felser}}]{Jungwirth2018}%
  \BibitemOpen
  \bibfield  {author} {\bibinfo {author} {\bibfnamefont {T.}~\bibnamefont
  {Jungwirth}}, \bibinfo {author} {\bibfnamefont {J.}~\bibnamefont {Sinova}},
  \bibinfo {author} {\bibfnamefont {A.}~\bibnamefont {Manchon}}, \bibinfo
  {author} {\bibfnamefont {X.}~\bibnamefont {Marti}}, \bibinfo {author}
  {\bibfnamefont {J.}~\bibnamefont {Wunderlich}},\ and\ \bibinfo {author}
  {\bibfnamefont {C.}~\bibnamefont {Felser}},\ }\href
  {https://doi.org/10.1038/s41567-018-0063-6} {\bibfield  {journal} {\bibinfo
  {journal} {Nature Physics}\ }\textbf {\bibinfo {volume} {14}},\ \bibinfo
  {pages} {200} (\bibinfo {year} {2018})}\BibitemShut {NoStop}%
\bibitem [{\citenamefont {Baltz}\ \emph {et~al.}(2018)\citenamefont {Baltz},
  \citenamefont {Manchon}, \citenamefont {Tsoi}, \citenamefont {Moriyama},
  \citenamefont {Ono},\ and\ \citenamefont {Tserkovnyak}}]{Baltz2018}%
  \BibitemOpen
  \bibfield  {author} {\bibinfo {author} {\bibfnamefont {V.}~\bibnamefont
  {Baltz}}, \bibinfo {author} {\bibfnamefont {A.}~\bibnamefont {Manchon}},
  \bibinfo {author} {\bibfnamefont {M.}~\bibnamefont {Tsoi}}, \bibinfo {author}
  {\bibfnamefont {T.}~\bibnamefont {Moriyama}}, \bibinfo {author}
  {\bibfnamefont {T.}~\bibnamefont {Ono}},\ and\ \bibinfo {author}
  {\bibfnamefont {Y.}~\bibnamefont {Tserkovnyak}},\ }\href
  {https://doi.org/10.1103/revmodphys.90.015005} {\bibfield  {journal}
  {\bibinfo  {journal} {Reviews of Modern Physics}\ }\textbf {\bibinfo {volume}
  {90}},\ \bibinfo {pages} {015005} (\bibinfo {year} {2018})}\BibitemShut
  {NoStop}%
\bibitem [{\citenamefont {Sasaki}\ \emph {et~al.}(1985)\citenamefont {Sasaki},
  \citenamefont {Yui},\ and\ \citenamefont {Yamaguchi}}]{Sasaki1985}%
  \BibitemOpen
  \bibfield  {author} {\bibinfo {author} {\bibfnamefont {A.}~\bibnamefont
  {Sasaki}}, \bibinfo {author} {\bibfnamefont {S.}~\bibnamefont {Yui}},\ and\
  \bibinfo {author} {\bibfnamefont {M.}~\bibnamefont {Yamaguchi}},\ }\href
  {https://doi.org/10.2465/minerj.12.393} {\bibfield  {journal} {\bibinfo
  {journal} {Mineral. J.}\ }\textbf {\bibinfo {volume} {12}},\ \bibinfo {pages}
  {393} (\bibinfo {year} {1985})}\BibitemShut {NoStop}%
\bibitem [{\citenamefont {Bertrand}\ and\ \citenamefont
  {Kerner-Czeskleba}(1975)}]{Bertrand1975}%
  \BibitemOpen
  \bibfield  {author} {\bibinfo {author} {\bibfnamefont {D.}~\bibnamefont
  {Bertrand}}\ and\ \bibinfo {author} {\bibfnamefont {H.}~\bibnamefont
  {Kerner-Czeskleba}},\ }\href
  {https://doi.org/10.1051/jphys:01975003605037900} {\bibfield  {journal}
  {\bibinfo  {journal} {J. Phys.}\ }\textbf {\bibinfo {volume} {36}},\ \bibinfo
  {pages} {379} (\bibinfo {year} {1975})}\BibitemShut {NoStop}%
\bibitem [{\citenamefont {McCarroll}\ \emph {et~al.}(1957)\citenamefont
  {McCarroll}, \citenamefont {Katz},\ and\ \citenamefont
  {Ward}}]{McCarroll1957}%
  \BibitemOpen
  \bibfield  {author} {\bibinfo {author} {\bibfnamefont {W.~H.}\ \bibnamefont
  {McCarroll}}, \bibinfo {author} {\bibfnamefont {L.}~\bibnamefont {Katz}},\
  and\ \bibinfo {author} {\bibfnamefont {R.}~\bibnamefont {Ward}},\ }\href
  {https://doi.org/https://doi.org/10.1021/ja01577a021} {\bibfield  {journal}
  {\bibinfo  {journal} {J. Am. Chem. Soc.}\ }\textbf {\bibinfo {volume} {79}},\
  \bibinfo {pages} {5410} (\bibinfo {year} {1957})}\BibitemShut {NoStop}%
\bibitem [{\citenamefont {McAlister}\ and\ \citenamefont
  {Strobel}(1983)}]{McAlister1983}%
  \BibitemOpen
  \bibfield  {author} {\bibinfo {author} {\bibfnamefont {S.~P.}\ \bibnamefont
  {McAlister}}\ and\ \bibinfo {author} {\bibfnamefont {P.}~\bibnamefont
  {Strobel}},\ }\href {https://doi.org/10.1016/0304-8853(83)90073-2} {\bibfield
   {journal} {\bibinfo  {journal} {J. Magn. Magn. Mater.}\ }\textbf {\bibinfo
  {volume} {30}},\ \bibinfo {pages} {340} (\bibinfo {year} {1983})}\BibitemShut
  {NoStop}%
\bibitem [{\citenamefont {Wang}\ \emph {et~al.}(2015)\citenamefont {Wang},
  \citenamefont {Pascut}, \citenamefont {Gao}, \citenamefont {Tyson},
  \citenamefont {Haule}, \citenamefont {Kiryukhin},\ and\ \citenamefont
  {Cheong}}]{Wang2015}%
  \BibitemOpen
  \bibfield  {author} {\bibinfo {author} {\bibfnamefont {Y.}~\bibnamefont
  {Wang}}, \bibinfo {author} {\bibfnamefont {G.~L.}\ \bibnamefont {Pascut}},
  \bibinfo {author} {\bibfnamefont {B.}~\bibnamefont {Gao}}, \bibinfo {author}
  {\bibfnamefont {T.~A.}\ \bibnamefont {Tyson}}, \bibinfo {author}
  {\bibfnamefont {K.}~\bibnamefont {Haule}}, \bibinfo {author} {\bibfnamefont
  {V.}~\bibnamefont {Kiryukhin}},\ and\ \bibinfo {author} {\bibfnamefont
  {S.-W.}\ \bibnamefont {Cheong}},\ }\href {https://doi.org/10.1038/srep12268}
  {\bibfield  {journal} {\bibinfo  {journal} {Sci. Rep.}\ }\textbf {\bibinfo
  {volume} {5}},\ \bibinfo {pages} {12268} (\bibinfo {year}
  {2015})}\BibitemShut {NoStop}%
\bibitem [{\citenamefont {Varret}\ \emph {et~al.}(1972)\citenamefont {Varret},
  \citenamefont {Kerner-Czeskleba}, \citenamefont {Hartmann-Boutron},\ and\
  \citenamefont {Imbert}}]{Varret1972}%
  \BibitemOpen
  \bibfield  {author} {\bibinfo {author} {\bibfnamefont {F.}~\bibnamefont
  {Varret}}, \bibinfo {author} {\bibfnamefont {H.}~\bibnamefont
  {Kerner-Czeskleba}}, \bibinfo {author} {\bibfnamefont {F.}~\bibnamefont
  {Hartmann-Boutron}},\ and\ \bibinfo {author} {\bibfnamefont {P.}~\bibnamefont
  {Imbert}},\ }\href {https://doi.org/10.1051/jphys:01972003305-6054900}
  {\bibfield  {journal} {\bibinfo  {journal} {J. Phys.}\ }\textbf {\bibinfo
  {volume} {3}},\ \bibinfo {pages} {549} (\bibinfo {year} {1972})}\BibitemShut
  {NoStop}%
\bibitem [{\citenamefont {Cotton}(1964)}]{Cotton1964}%
  \BibitemOpen
  \bibfield  {author} {\bibinfo {author} {\bibfnamefont {F.~A.}\ \bibnamefont
  {Cotton}},\ }\href {https://doi.org/10.1021/ic50019a003} {\bibfield
  {journal} {\bibinfo  {journal} {Inorg. Chem.}\ }\textbf {\bibinfo {volume}
  {3}},\ \bibinfo {pages} {1217} (\bibinfo {year} {1964})}\BibitemShut
  {NoStop}%
\bibitem [{\citenamefont {Kurumaji}\ \emph {et~al.}(2015)\citenamefont
  {Kurumaji}, \citenamefont {Ishiwata},\ and\ \citenamefont
  {Tokura}}]{Kurumaji2015}%
  \BibitemOpen
  \bibfield  {author} {\bibinfo {author} {\bibfnamefont {T.}~\bibnamefont
  {Kurumaji}}, \bibinfo {author} {\bibfnamefont {S.}~\bibnamefont {Ishiwata}},\
  and\ \bibinfo {author} {\bibfnamefont {Y.}~\bibnamefont {Tokura}},\ }\href
  {https://doi.org/10.1103/PhysRevX.5.031034} {\bibfield  {journal} {\bibinfo
  {journal} {Phys. Rev. X}\ }\textbf {\bibinfo {volume} {5}},\ \bibinfo {pages}
  {031034} (\bibinfo {year} {2015})}\BibitemShut {NoStop}%
\bibitem [{\citenamefont {Kurumaji}\ \emph
  {et~al.}(2017{\natexlab{a}})\citenamefont {Kurumaji}, \citenamefont
  {Takahashi}, \citenamefont {Fujioka}, \citenamefont {Masuda}, \citenamefont
  {Shishikura}, \citenamefont {Ishiwata},\ and\ \citenamefont
  {Tokura}}]{Kurumaji2017a}%
  \BibitemOpen
  \bibfield  {author} {\bibinfo {author} {\bibfnamefont {T.}~\bibnamefont
  {Kurumaji}}, \bibinfo {author} {\bibfnamefont {Y.}~\bibnamefont {Takahashi}},
  \bibinfo {author} {\bibfnamefont {J.}~\bibnamefont {Fujioka}}, \bibinfo
  {author} {\bibfnamefont {R.}~\bibnamefont {Masuda}}, \bibinfo {author}
  {\bibfnamefont {H.}~\bibnamefont {Shishikura}}, \bibinfo {author}
  {\bibfnamefont {S.}~\bibnamefont {Ishiwata}},\ and\ \bibinfo {author}
  {\bibfnamefont {Y.}~\bibnamefont {Tokura}},\ }\href
  {https://doi.org/10.1103/PhysRevB.95.020405 Cross} {\bibfield  {journal}
  {\bibinfo  {journal} {Phys. Rev. B}\ }\textbf {\bibinfo {volume} {95}},\
  \bibinfo {pages} {020405(R)} (\bibinfo {year}
  {2017}{\natexlab{a}})}\BibitemShut {NoStop}%
\bibitem [{\citenamefont {Kurumaji}\ \emph
  {et~al.}(2017{\natexlab{b}})\citenamefont {Kurumaji}, \citenamefont
  {Takahashi}, \citenamefont {Fujioka}, \citenamefont {Masuda}, \citenamefont
  {Shishikura}, \citenamefont {Ishiwata},\ and\ \citenamefont
  {Tokura}}]{Kurumaji2017}%
  \BibitemOpen
  \bibfield  {author} {\bibinfo {author} {\bibfnamefont {T.}~\bibnamefont
  {Kurumaji}}, \bibinfo {author} {\bibfnamefont {Y.}~\bibnamefont {Takahashi}},
  \bibinfo {author} {\bibfnamefont {J.}~\bibnamefont {Fujioka}}, \bibinfo
  {author} {\bibfnamefont {R.}~\bibnamefont {Masuda}}, \bibinfo {author}
  {\bibfnamefont {H.}~\bibnamefont {Shishikura}}, \bibinfo {author}
  {\bibfnamefont {S.}~\bibnamefont {Ishiwata}},\ and\ \bibinfo {author}
  {\bibfnamefont {Y.}~\bibnamefont {Tokura}},\ }\href
  {https://doi.org/10.1103/PhysRevLett.119.077206} {\bibfield  {journal}
  {\bibinfo  {journal} {Phys. Rev. Lett.}\ }\textbf {\bibinfo {volume} {119}},\
  \bibinfo {pages} {077206} (\bibinfo {year} {2017}{\natexlab{b}})}\BibitemShut
  {NoStop}%
\bibitem [{\citenamefont {Yu}\ \emph {et~al.}(2018)\citenamefont {Yu},
  \citenamefont {Gao}, \citenamefont {Kim}, \citenamefont {Cheong},
  \citenamefont {Man}, \citenamefont {Mad{\'{e}}o}, \citenamefont {Dani},\ and\
  \citenamefont {Talbayev}}]{Yu2018}%
  \BibitemOpen
  \bibfield  {author} {\bibinfo {author} {\bibfnamefont {S.}~\bibnamefont
  {Yu}}, \bibinfo {author} {\bibfnamefont {B.}~\bibnamefont {Gao}}, \bibinfo
  {author} {\bibfnamefont {J.~W.}\ \bibnamefont {Kim}}, \bibinfo {author}
  {\bibfnamefont {S.~W.}\ \bibnamefont {Cheong}}, \bibinfo {author}
  {\bibfnamefont {M.~K.~L.}\ \bibnamefont {Man}}, \bibinfo {author} {\bibfnamefont
  {J.}~\bibnamefont {Madeo}}, \bibinfo {author} {\bibfnamefont {K.~M.}\
  \bibnamefont {Dani}},\ and\ \bibinfo {author} {\bibfnamefont
  {D.}~\bibnamefont {Talbayev}},\ }\href
  {https://doi.org/10.1103/PhysRevLett.120.037601} {\bibfield  {journal}
  {\bibinfo  {journal} {Phys. Rev. Lett.}\ }\textbf {\bibinfo {volume} {120}},\
  \bibinfo {pages} {037601} (\bibinfo {year} {2018})}\BibitemShut {NoStop}%
\bibitem [{\citenamefont {Solovyev}\ and\ \citenamefont
  {Streltsov}(2019)}]{Solovyev2019}%
  \BibitemOpen
  \bibfield  {author} {\bibinfo {author} {\bibfnamefont {I.~V.}\ \bibnamefont
  {Solovyev}}\ and\ \bibinfo {author} {\bibfnamefont {S.~V.}\ \bibnamefont
  {Streltsov}},\ }\href {https://doi.org/10.1103/PhysRevMaterials.3.114402}
  {\bibfield  {journal} {\bibinfo  {journal} {Phys. Rev. Mater.}\ }\textbf
  {\bibinfo {volume} {3}},\ \bibinfo {pages} {114402} (\bibinfo {year}
  {2019})}\BibitemShut {NoStop}%
\bibitem [{\citenamefont {Nakayama}\ \emph {et~al.}(2011)\citenamefont
  {Nakayama}, \citenamefont {Nakamura},\ and\ \citenamefont
  {Akaki}}]{Nakayama2011}%
  \BibitemOpen
  \bibfield  {author} {\bibinfo {author} {\bibfnamefont {S.}~\bibnamefont
  {Nakayama}}, \bibinfo {author} {\bibfnamefont {R.}~\bibnamefont {Nakamura}},\
  and\ \bibinfo {author} {\bibfnamefont {M.}~\bibnamefont {Akaki}},\ }\href
  {https://doi.org/10.1143/JPSJ.80.104706} {\bibfield  {journal} {\bibinfo
  {journal} {J. Phys. Soc. Japan}\ }\textbf {\bibinfo {volume} {80}},\ \bibinfo
  {pages} {104706} (\bibinfo {year} {2011})}\BibitemShut {NoStop}%
\bibitem [{\citenamefont {Streltsov}\ \emph {et~al.}(2019)\citenamefont
  {Streltsov}, \citenamefont {Huang}, \citenamefont {Solovyev},\ and\
  \citenamefont {Khomskii}}]{Streltsov2019}%
  \BibitemOpen
  \bibfield  {author} {\bibinfo {author} {\bibfnamefont {S.~V.}\ \bibnamefont
  {Streltsov}}, \bibinfo {author} {\bibfnamefont {D.-J.}\ \bibnamefont
  {Huang}}, \bibinfo {author} {\bibfnamefont {I.}~\bibnamefont {Solovyev}},\
  and\ \bibinfo {author} {\bibfnamefont {D.~I.}\ \bibnamefont {Khomskii}},\
  }\href {https://doi.org/10.1134/S0021364019120026} {\bibfield  {journal}
  {\bibinfo  {journal} {JETP Lett.}\ }\textbf {\bibinfo {volume} {109}},\
  \bibinfo {pages} {786} (\bibinfo {year} {2019})}\BibitemShut {NoStop}%
\bibitem [{\citenamefont {Bertinshaw}\ \emph {et~al.}(2014)\citenamefont
  {Bertinshaw}, \citenamefont {Ulrich}, \citenamefont {G{\"{u}}nther},
  \citenamefont {Schrettle}, \citenamefont {Wohlauer}, \citenamefont {Krohns},
  \citenamefont {Reehuis}, \citenamefont {Studer}, \citenamefont {Avdeev},
  \citenamefont {Quach}, \citenamefont {Groza}, \citenamefont {Tsurkan},
  \citenamefont {Loidl},\ and\ \citenamefont {Deisenhofer}}]{Bertinshaw2014}%
  \BibitemOpen
  \bibfield  {author} {\bibinfo {author} {\bibfnamefont {J.}~\bibnamefont
  {Bertinshaw}}, \bibinfo {author} {\bibfnamefont {C.}~\bibnamefont {Ulrich}},
  \bibinfo {author} {\bibfnamefont {A.}~\bibnamefont {G{\"{u}}nther}}, \bibinfo
  {author} {\bibfnamefont {F.}~\bibnamefont {Schrettle}}, \bibinfo {author}
  {\bibfnamefont {M.}~\bibnamefont {Wohlauer}}, \bibinfo {author}
  {\bibfnamefont {S.}~\bibnamefont {Krohns}}, \bibinfo {author} {\bibfnamefont
  {M.}~\bibnamefont {Reehuis}}, \bibinfo {author} {\bibfnamefont {A.~J.}\
  \bibnamefont {Studer}}, \bibinfo {author} {\bibfnamefont {M.}~\bibnamefont
  {Avdeev}}, \bibinfo {author} {\bibfnamefont {D.~V.}\ \bibnamefont {Quach}},
  \bibinfo {author} {\bibfnamefont {J.~R.}\ \bibnamefont {Groza}}, \bibinfo
  {author} {\bibfnamefont {V.}~\bibnamefont {Tsurkan}}, \bibinfo {author}
  {\bibfnamefont {A.}~\bibnamefont {Loidl}},\ and\ \bibinfo {author}
  {\bibfnamefont {J.}~\bibnamefont {Deisenhofer}},\ }\href
  {https://doi.org/10.1038/srep06079} {\bibfield  {journal} {\bibinfo
  {journal} {Sci. Rep.}\ }\textbf {\bibinfo {volume} {4}},\ \bibinfo {pages}
  {6078} (\bibinfo {year} {2014})}\BibitemShut {NoStop}%
\bibitem [{\citenamefont {Abragam}\ and\ \citenamefont
  {Bleaney}(1970)}]{Abragam1970}%
  \BibitemOpen
  \bibfield  {author} {\bibinfo {author} {\bibfnamefont {A.}~\bibnamefont
  {Abragam}}\ and\ \bibinfo {author} {\bibfnamefont {B.}~\bibnamefont
  {Bleaney}},\ }\href@noop {} {\emph {\bibinfo {title} {{Electron Paramagnetic
  Resonance of Transition Ions}}}}\ (\bibinfo  {publisher} {Oxford University
  Press},\ \bibinfo {address} {Oxford},\ \bibinfo {year} {1970})\BibitemShut
  {NoStop}%
\bibitem [{\citenamefont {Reschke}\ \emph {et~al.}(2020)\citenamefont
  {Reschke}, \citenamefont {Tsirlin}, \citenamefont {Khan}, \citenamefont
  {Prodan}, \citenamefont {Tsurkan}, \citenamefont {Kézsmárki},\ and\
  \citenamefont {Deisenhofer}}]{Reschke2020}%
  \BibitemOpen
  \bibfield  {author} {\bibinfo {author} {\bibfnamefont {S.}~\bibnamefont
  {Reschke}}, \bibinfo {author} {\bibfnamefont {A.}~\bibnamefont {Tsirlin}},
  \bibinfo {author} {\bibfnamefont {N.}~\bibnamefont {Khan}}, \bibinfo {author}
  {\bibfnamefont {L.}~\bibnamefont {Prodan}}, \bibinfo {author} {\bibfnamefont
  {V.}~\bibnamefont {Tsurkan}}, \bibinfo {author} {\bibfnamefont
  {I.}~\bibnamefont {Kézsmárki}},\ and\ \bibinfo {author} {\bibfnamefont
  {J.}~\bibnamefont {Deisenhofer}},\ }\href
  {https://doi.org/10.1103/PhysRevB.102.094307} {\bibfield  {journal} {\bibinfo
   {journal} {Physical Review B}\ }\textbf {\bibinfo {volume} {102}},\ \bibinfo
  {pages} {094307} (\bibinfo {year} {2020})}\BibitemShut {NoStop}%
\bibitem [{\citenamefont {Stanislavchuk}\ \emph {et~al.}(2020)\citenamefont
  {Stanislavchuk}, \citenamefont {Pascut}, \citenamefont {Litvinchuk},
  \citenamefont {Liu}, \citenamefont {Choi}, \citenamefont {Gutmann},\ and\
  \citenamefont {Gao}}]{Stanislavchuk2020}%
  \BibitemOpen
  \bibfield  {author} {\bibinfo {author} {\bibfnamefont {T.~N.}\ \bibnamefont
  {Stanislavchuk}}, \bibinfo {author} {\bibfnamefont {G.~L.}\ \bibnamefont
  {Pascut}}, \bibinfo {author} {\bibfnamefont {A.~P.}\ \bibnamefont
  {Litvinchuk}}, \bibinfo {author} {\bibfnamefont {Z.}~\bibnamefont {Liu}},
  \bibinfo {author} {\bibfnamefont {S.}~\bibnamefont {Choi}}, \bibinfo {author}
  {\bibfnamefont {M.~J.}\ \bibnamefont {Gutmann}},\ and\ \bibinfo {author}
  {\bibfnamefont {B.}~\bibnamefont {Gao}},\ }\href
  {https://doi.org/10.1103/PhysRevB.102.115139} {\bibfield  {journal} {\bibinfo
   {journal} {Physical Review B}\ }\textbf {\bibinfo {volume} {102}},\ \bibinfo
  {pages} {115139} (\bibinfo {year} {2020})}\BibitemShut {NoStop}%
\bibitem [{Str()}]{Strinic2020}%
  \BibitemOpen
  \href@noop {} {\bibinfo {title} {{A. Strinic et al.
  (unpublished)}}}\BibitemShut {NoStop}%
\bibitem [{\citenamefont {Penc}\ \emph {et~al.}(2012)\citenamefont {Penc},
  \citenamefont {Romh{\'{a}}nyi}, \citenamefont {R{\~{o}}{\~{o}}m},
  \citenamefont {Nagel}, \citenamefont {Antal}, \citenamefont {Feh{\'{e}}r},
  \citenamefont {J{\'{a}}nossy}, \citenamefont {Engelkamp}, \citenamefont
  {Murakawa}, \citenamefont {Tokura}, \citenamefont {Szaller}, \citenamefont
  {Bord{\'{a}}cs},\ and\ \citenamefont {K{\'{e}}zsm{\'{a}}rki}}]{Penc2012}%
  \BibitemOpen
  \bibfield  {author} {\bibinfo {author} {\bibfnamefont {K.}~\bibnamefont
  {Penc}}, \bibinfo {author} {\bibfnamefont {J.}~\bibnamefont
  {Romh{\'{a}}nyi}}, \bibinfo {author} {\bibfnamefont {T.}~\bibnamefont
  {R{\~{o}}{\~{o}}m}}, \bibinfo {author} {\bibfnamefont {U.}~\bibnamefont
  {Nagel}}, \bibinfo {author} {\bibfnamefont {A.}~\bibnamefont {Antal}},
  \bibinfo {author} {\bibfnamefont {T.}~\bibnamefont {Feh{\'{e}}r}}, \bibinfo
  {author} {\bibfnamefont {A.}~\bibnamefont {J{\'{a}}nossy}}, \bibinfo {author}
  {\bibfnamefont {H.}~\bibnamefont {Engelkamp}}, \bibinfo {author}
  {\bibfnamefont {H.}~\bibnamefont {Murakawa}}, \bibinfo {author}
  {\bibfnamefont {Y.}~\bibnamefont {Tokura}}, \bibinfo {author} {\bibfnamefont
  {D.}~\bibnamefont {Szaller}}, \bibinfo {author} {\bibfnamefont
  {S.}~\bibnamefont {Bord{\'{a}}cs}},\ and\ \bibinfo {author} {\bibfnamefont
  {I.}~\bibnamefont {K{\'{e}}zsm{\'{a}}rki}},\ }\href
  {https://doi.org/10.1103/PhysRevLett.108.257203} {\bibfield  {journal}
  {\bibinfo  {journal} {Phys. Rev. Lett.}\ }\textbf {\bibinfo {volume} {108}},\
  \bibinfo {pages} {257203} (\bibinfo {year} {2012})}\BibitemShut {NoStop}%
\bibitem [{\citenamefont {Kocsis}\ \emph {et~al.}(2018)\citenamefont {Kocsis},
  \citenamefont {Penc}, \citenamefont {R{\~{o}}{\~{o}}m}, \citenamefont
  {Nagel}, \citenamefont {V{\'{i}}t}, \citenamefont {Romh{\'{a}}nyi},
  \citenamefont {Tokunaga}, \citenamefont {Taguchi}, \citenamefont {Tokura},
  \citenamefont {K{\'{e}}zsm{\'{a}}rki},\ and\ \citenamefont
  {Bord{\'{a}}cs}}]{Kocsis2018a}%
  \BibitemOpen
  \bibfield  {author} {\bibinfo {author} {\bibfnamefont {V.}~\bibnamefont
  {Kocsis}}, \bibinfo {author} {\bibfnamefont {K.}~\bibnamefont {Penc}},
  \bibinfo {author} {\bibfnamefont {T.}~\bibnamefont {R{\~{o}}{\~{o}}m}},
  \bibinfo {author} {\bibfnamefont {U.}~\bibnamefont {Nagel}}, \bibinfo
  {author} {\bibfnamefont {J.}~\bibnamefont {V{\'{i}}t}}, \bibinfo {author}
  {\bibfnamefont {J.}~\bibnamefont {Romh{\'{a}}nyi}}, \bibinfo {author}
  {\bibfnamefont {Y.}~\bibnamefont {Tokunaga}}, \bibinfo {author}
  {\bibfnamefont {Y.}~\bibnamefont {Taguchi}}, \bibinfo {author} {\bibfnamefont
  {Y.}~\bibnamefont {Tokura}}, \bibinfo {author} {\bibfnamefont
  {I.}~\bibnamefont {K{\'{e}}zsm{\'{a}}rki}},\ and\ \bibinfo {author}
  {\bibfnamefont {S.}~\bibnamefont {Bord{\'{a}}cs}},\ }\href
  {https://doi.org/10.1103/PhysRevLett.121.057601} {\bibfield  {journal}
  {\bibinfo  {journal} {Phys. Rev. Lett.}\ }\textbf {\bibinfo {volume} {121}},\
  \bibinfo {pages} {057601} (\bibinfo {year} {2018})}\BibitemShut {NoStop}%
\bibitem [{\citenamefont {Osborn}(1945)}]{Osborn1945}%
  \BibitemOpen
  \bibfield  {author} {\bibinfo {author} {\bibfnamefont {J.~A.}\ \bibnamefont
  {Osborn}},\ }\href {https://doi.org/10.1103/physrev.67.351} {\bibfield
  {journal} {\bibinfo  {journal} {Physical Review}\ }\textbf {\bibinfo {volume}
  {67}},\ \bibinfo {pages} {351} (\bibinfo {year} {1945})}\BibitemShut
  {NoStop}%
\bibitem [{\citenamefont {Testelin}\ \emph {et~al.}(1992)\citenamefont
  {Testelin}, \citenamefont {Rigaux}, \citenamefont {Mauger}, \citenamefont
  {Mycielski},\ and\ \citenamefont {Julien}}]{Testelin1992}%
  \BibitemOpen
  \bibfield  {author} {\bibinfo {author} {\bibfnamefont {C.}~\bibnamefont
  {Testelin}}, \bibinfo {author} {\bibfnamefont {C.}~\bibnamefont {Rigaux}},
  \bibinfo {author} {\bibfnamefont {A.}~\bibnamefont {Mauger}}, \bibinfo
  {author} {\bibfnamefont {A.}~\bibnamefont {Mycielski}},\ and\ \bibinfo
  {author} {\bibfnamefont {C.}~\bibnamefont {Julien}},\ }\href
  {https://doi.org/10.1103/physrevb.46.2183} {\bibfield  {journal} {\bibinfo
  {journal} {Physical Review B}\ }\textbf {\bibinfo {volume} {46}},\ \bibinfo
  {pages} {2183} (\bibinfo {year} {1992})}\BibitemShut {NoStop}%
\bibitem [{\citenamefont {Mittelstädt}\ \emph {et~al.}(2015)\citenamefont
  {Mittelstädt}, \citenamefont {Schmidt}, \citenamefont {Wang}, \citenamefont
  {Mayr}, \citenamefont {Tsurkan}, \citenamefont {Lunkenheimer}, \citenamefont
  {Ish}, \citenamefont {Balents}, \citenamefont {Deisenhofer},\ and\
  \citenamefont {Loidl}}]{Mittelstaedt2015}%
  \BibitemOpen
  \bibfield  {author} {\bibinfo {author} {\bibfnamefont {L.}~\bibnamefont
  {Mittelstädt}}, \bibinfo {author} {\bibfnamefont {M.}~\bibnamefont
  {Schmidt}}, \bibinfo {author} {\bibfnamefont {Z.}~\bibnamefont {Wang}},
  \bibinfo {author} {\bibfnamefont {F.}~\bibnamefont {Mayr}}, \bibinfo {author}
  {\bibfnamefont {V.}~\bibnamefont {Tsurkan}}, \bibinfo {author} {\bibfnamefont
  {P.}~\bibnamefont {Lunkenheimer}}, \bibinfo {author} {\bibfnamefont
  {D.}~\bibnamefont {Ish}}, \bibinfo {author} {\bibfnamefont {L.}~\bibnamefont
  {Balents}}, \bibinfo {author} {\bibfnamefont {J.}~\bibnamefont
  {Deisenhofer}},\ and\ \bibinfo {author} {\bibfnamefont {A.}~\bibnamefont
  {Loidl}},\ }\href {https://doi.org/10.1103/PhysRevB.91.125112} {\bibfield
  {journal} {\bibinfo  {journal} {Physical Review B}\ }\textbf {\bibinfo
  {volume} {91}},\ \bibinfo {pages} {125112} (\bibinfo {year}
  {2015})}\BibitemShut {NoStop}%
\bibitem [{\citenamefont {Laurita}\ \emph {et~al.}(2015)\citenamefont
  {Laurita}, \citenamefont {Deisenhofer}, \citenamefont {Pan}, \citenamefont
  {Morris}, \citenamefont {Schmidt}, \citenamefont {Johnsson}, \citenamefont
  {Tsurkan}, \citenamefont {Loidl},\ and\ \citenamefont
  {Armitage}}]{Laurita2015a}%
  \BibitemOpen
  \bibfield  {author} {\bibinfo {author} {\bibfnamefont {N.}~\bibnamefont
  {Laurita}}, \bibinfo {author} {\bibfnamefont {J.}~\bibnamefont
  {Deisenhofer}}, \bibinfo {author} {\bibfnamefont {L.}~\bibnamefont {Pan}},
  \bibinfo {author} {\bibfnamefont {C.}~\bibnamefont {Morris}}, \bibinfo
  {author} {\bibfnamefont {M.}~\bibnamefont {Schmidt}}, \bibinfo {author}
  {\bibfnamefont {M.}~\bibnamefont {Johnsson}}, \bibinfo {author}
  {\bibfnamefont {V.}~\bibnamefont {Tsurkan}}, \bibinfo {author} {\bibfnamefont
  {A.}~\bibnamefont {Loidl}},\ and\ \bibinfo {author} {\bibfnamefont
  {N.}~\bibnamefont {Armitage}},\ }\href
  {https://doi.org/10.1103/physrevlett.114.207201} {\bibfield  {journal}
  {\bibinfo  {journal} {Physical Review Letters}\ }\textbf {\bibinfo {volume}
  {114}},\ \bibinfo {pages} {207201} (\bibinfo {year} {2015})}\BibitemShut
  {NoStop}%
\bibitem [{\citenamefont {Strinic}\ \emph {et~al.}()\citenamefont {Strinic},
  \citenamefont {Reschke}, \citenamefont {Vasin}, \citenamefont {Schmidt},
  \citenamefont {Loidl}, \citenamefont {Tsurkan}, \citenamefont {Eremin},\ and\
  \citenamefont {Deisenhofer}}]{Strinic2020a}%
  \BibitemOpen
  \bibfield  {author} {\bibinfo {author} {\bibfnamefont {A.}~\bibnamefont
  {Strinic}}, \bibinfo {author} {\bibfnamefont {S.}~\bibnamefont {Reschke}},
  \bibinfo {author} {\bibfnamefont {K.~V.}\ \bibnamefont {Vasin}}, \bibinfo
  {author} {\bibfnamefont {M.}~\bibnamefont {Schmidt}}, \bibinfo {author}
  {\bibfnamefont {A.}~\bibnamefont {Loidl}}, \bibinfo {author} {\bibfnamefont
  {V.}~\bibnamefont {Tsurkan}}, \bibinfo {author} {\bibfnamefont {M.~V.}\
  \bibnamefont {Eremin}},\ and\ \bibinfo {author} {\bibfnamefont
  {J.}~\bibnamefont {Deisenhofer}},\ }\href
  {https://doi.org/10.1103/PhysRevB.102.134409} {\bibfield  {journal} {\bibinfo
  {journal} {Physical Review B}\ }\textbf {\bibinfo {volume} {102}},\ \bibinfo
  {pages} {134409} (\bibinfo {year} {2020})}\BibitemShut {NoStop}%
\bibitem [{\citenamefont {Mai}\ \emph {et~al.}(2016)\citenamefont {Mai},
  \citenamefont {Svoboda}, \citenamefont {Warren}, \citenamefont {Jang},
  \citenamefont {Brangham}, \citenamefont {Jeong}, \citenamefont {Cheong},\
  and\ \citenamefont {Aguilar}}]{Mai2016}%
  \BibitemOpen
  \bibfield  {author} {\bibinfo {author} {\bibfnamefont {T.~T.}\ \bibnamefont
  {Mai}}, \bibinfo {author} {\bibfnamefont {C.}~\bibnamefont {Svoboda}},
  \bibinfo {author} {\bibfnamefont {M.~T.}\ \bibnamefont {Warren}}, \bibinfo
  {author} {\bibfnamefont {T.-H.}\ \bibnamefont {Jang}}, \bibinfo {author}
  {\bibfnamefont {J.}~\bibnamefont {Brangham}}, \bibinfo {author}
  {\bibfnamefont {Y.~H.}\ \bibnamefont {Jeong}}, \bibinfo {author}
  {\bibfnamefont {S.-W.}\ \bibnamefont {Cheong}},\ and\ \bibinfo {author}
  {\bibfnamefont {R.~V.}\ \bibnamefont {Aguilar}},\ }\href
  {https://doi.org/10.1103/physrevb.94.224416} {\bibfield  {journal} {\bibinfo
  {journal} {Physical Review B}\ }\textbf {\bibinfo {volume} {94}},\ \bibinfo
  {pages} {224416} (\bibinfo {year} {2016})}\BibitemShut {NoStop}%
\bibitem [{\citenamefont {Vallin}(1970)}]{Vallin1970}%
  \BibitemOpen
  \bibfield  {author} {\bibinfo {author} {\bibfnamefont {J.~T.}\ \bibnamefont
  {Vallin}},\ }\href {https://doi.org/10.1103/PhysRevB.2.2390} {\bibfield
  {journal} {\bibinfo  {journal} {Phys. Rev. B}\ }\textbf {\bibinfo {volume}
  {2}},\ \bibinfo {pages} {2390} (\bibinfo {year} {1970})}\BibitemShut
  {NoStop}%
\bibitem [{\citenamefont {Vogel}\ and\ \citenamefont
  {Rivera-Iratchet}(1980)}]{Vogel1980}%
  \BibitemOpen
  \bibfield  {author} {\bibinfo {author} {\bibfnamefont {E.~E.}\ \bibnamefont
  {Vogel}}\ and\ \bibinfo {author} {\bibfnamefont {J.}~\bibnamefont
  {Rivera-Iratchet}},\ }\href {https://doi.org/10.1103/PhysRevB.22.4511}
  {\bibfield  {journal} {\bibinfo  {journal} {Phys. Rev. B}\ }\textbf {\bibinfo
  {volume} {22}},\ \bibinfo {pages} {4511} (\bibinfo {year}
  {1980})}\BibitemShut {NoStop}%
\bibitem [{\citenamefont {Wittekoek}\ \emph {et~al.}(1973)\citenamefont
  {Wittekoek}, \citenamefont {{Van Stapele}},\ and\ \citenamefont
  {Wijma}}]{Wittekoek1973}%
  \BibitemOpen
  \bibfield  {author} {\bibinfo {author} {\bibfnamefont {S.}~\bibnamefont
  {Wittekoek}}, \bibinfo {author} {\bibfnamefont {R.~P.}\ \bibnamefont {{Van
  Stapele}}},\ and\ \bibinfo {author} {\bibfnamefont {A.~W.}\ \bibnamefont
  {Wijma}},\ }\href {https://doi.org/10.1103/PhysRevB.7.1667} {\bibfield
  {journal} {\bibinfo  {journal} {Phys. Rev. B}\ }\textbf {\bibinfo {volume}
  {7}},\ \bibinfo {pages} {1667} (\bibinfo {year} {1973})}\BibitemShut
  {NoStop}%
\bibitem [{\citenamefont {Feiner}(1982)}]{Feiner1982}%
  \BibitemOpen
  \bibfield  {author} {\bibinfo {author} {\bibfnamefont {L.~F.}\ \bibnamefont
  {Feiner}},\ }\href {https://doi.org/10.1088/0022-3719/15/7/017} {\bibfield
  {journal} {\bibinfo  {journal} {J. Phys. C Solid State Phys.}\ }\textbf
  {\bibinfo {volume} {15}},\ \bibinfo {pages} {1515} (\bibinfo {year}
  {1982})}\BibitemShut {NoStop}%
\bibitem [{\citenamefont {Stoneham}(1975)}]{Stoneham1975}%
  \BibitemOpen
  \bibfield  {author} {\bibinfo {author} {\bibfnamefont {A.~M.}\ \bibnamefont
  {Stoneham}},\ }\href@noop {} {\emph {\bibinfo {title} {{Theory of Defects in
  Solids}}}}\ (\bibinfo  {publisher} {Oxford University Press},\ \bibinfo
  {address} {Oxford},\ \bibinfo {year} {1975})\BibitemShut {NoStop}%
\bibitem [{\citenamefont {Toyozawa}(2003)}]{Toyozawa2003}%
  \BibitemOpen
  \bibfield  {author} {\bibinfo {author} {\bibfnamefont {Y.}~\bibnamefont
  {Toyozawa}},\ }\href@noop {} {\emph {\bibinfo {title} {{Optical Processes in
  Solids}}}}\ (\bibinfo  {publisher} {Cambridge University Press},\ \bibinfo
  {address} {New York},\ \bibinfo {year} {2003})\BibitemShut {NoStop}%
\bibitem [{\citenamefont {Pullerits}\ \emph {et~al.}(1995)\citenamefont
  {Pullerits}, \citenamefont {Monshouwer}, \citenamefont {van Mourik},\ and\
  \citenamefont {van Grondelle}}]{Pullerits1995}%
  \BibitemOpen
  \bibfield  {author} {\bibinfo {author} {\bibfnamefont {T.}~\bibnamefont
  {Pullerits}}, \bibinfo {author} {\bibfnamefont {R.}~\bibnamefont
  {Monshouwer}}, \bibinfo {author} {\bibfnamefont {F.}~\bibnamefont {van
  Mourik}},\ and\ \bibinfo {author} {\bibfnamefont {R.}~\bibnamefont {van
  Grondelle}},\ }\href {https://doi.org/10.1016/0301-0104(95)00073-W}
  {\bibfield  {journal} {\bibinfo  {journal} {Chem. Phys.}\ }\textbf {\bibinfo
  {volume} {194}},\ \bibinfo {pages} {395} (\bibinfo {year}
  {1995})}\BibitemShut {NoStop}%
\bibitem [{\citenamefont {R{\"{a}}tsep}\ \emph {et~al.}(2014)\citenamefont
  {R{\"{a}}tsep}, \citenamefont {Pajusalu}, \citenamefont {Linnanto},\ and\
  \citenamefont {Freiberg}}]{Ratsep2014}%
  \BibitemOpen
  \bibfield  {author} {\bibinfo {author} {\bibfnamefont {M.}~\bibnamefont
  {R{\"{a}}tsep}}, \bibinfo {author} {\bibfnamefont {M.}~\bibnamefont
  {Pajusalu}}, \bibinfo {author} {\bibfnamefont {J.~M.}\ \bibnamefont
  {Linnanto}},\ and\ \bibinfo {author} {\bibfnamefont {A.}~\bibnamefont
  {Freiberg}},\ }\href {http://aip.scitation.org/doi/10.1063/1.4897637}
  {\bibfield  {journal} {\bibinfo  {journal} {J. Chem. Phys.}\ }\textbf
  {\bibinfo {volume} {141}},\ \bibinfo {pages} {155102} (\bibinfo {year}
  {2014})}\BibitemShut {NoStop}%
\end{thebibliography}

%

\end{document}